\documentclass[fleqn,usenatbib]{mnras}

\usepackage{newtxtext,newtxmath}

\usepackage[T1]{fontenc}
\usepackage{ae,aecompl}

\usepackage{graphicx}	
\usepackage{amsmath}	

\def\beq{\begin{equation}}
\def\eeq{\end{equation}}
\def\beqa{\begin{eqnarray}}
\def\eeqa{\end{eqnarray}}
\def\wslap{\textsc{wslap+}}

\def\mnras{MNRAS}
\def\apj{ApJ}

\def\aap{A\&A}
\def\apjl{ApJ Letters}
\def\apjs{ApJS}

\title[Magnification in the \textit{Hubble Frontier Fields}]{Probability of magnification in the \textit{Hubble Frontier Fields} clusters}
\author[Vega-Ferrero, J. et al.]
{\parbox{\textwidth}{
J. Vega-Ferrero$^{1,2}$\thanks{E-mail: vegaf@sas.upenn.edu}, J. M. Diego$^{2}$, G. M. Bernstein$^{1}$}\\
$^1$ Department of Physics and Astronomy, University of Pennsylvania, 209 S. 33rd St, Philadelphia, PA 19104, USA\\
$^2$  IFCA, Instituto de F\'isica de Cantabria (UC-CSIC), Av. de Los Castros s/n, 39005 Santander, Spain}

\begin{document}

\date{Accepted ???. Received ???; in original form ???}

\pagerange{\pageref{firstpage}--\pageref{lastpage}} \pubyear{2018}

\maketitle
\label{firstpage}

\begin{abstract} 
We present free-form gravitational lensing models derived with the \wslap~code for the six \textit{Hubble Frontier Fields} clusters using the latest data available from the \textit{Frontier Fields Lensing Models v.4} collaboration. We present magnifications maps in the lens plane and caustic maps in the source plane. From these maps, we derive the probability of magnification using two different, but related, methods. We confirm MACS 0717 as the cluster with the most complex structure, and having the largest lensing efficiency and Einstein radius. When comparing these results with the ones obtained by previous observations of these clusters, we obtain an increase in the lensing efficiency between 1.4 and 2.3. We also find a good correlation with a relatively small dispersion between the lensing efficiency and Einstein radius as a function of the source redshift ($z_s$). Finally, we estimate the lensing effects produced by the six \textit{Hubble Frontier Fields} clusters on the luminosity function of galaxies at high redshift ($z=9$) for standard luminosity functions and an alternative luminosity function based on predictions from wave dark matter ($\psi$DM) models.
\end{abstract}


\section{Introduction}
\label{sec:intro}
Gravitational lensing has proven to be the most powerful tool to map the distribution of mass, in particular the invisible dark matter. Gravitational lensing is also used as a way of increasing the effective depth of observations by  taking advantage of the natural boost provided by magnification factors of a few to several tens. In the case of galaxy clusters, gravitational lensing can magnify a considerable area (tens of square arcseconds) at high redshift above significant magnification factors  ($\mu>10$). Deep observations, made for instance with the Hubble Space Telescope (HST), can detect faint sources that otherwise would be unobserved (at $\mu=10$ a source would gain 2.5 magnitudes). Looking for faint strongly magnified  galaxies has yielded some of the most distant galaxies ever observed \citep{Zheng2012,Coe2013,Zitrin2014,Bouwens2014,Infante2015,Hoag2017,Salmon2018}. More recently, a serendipitous search for SNe resulted in the discovery of the most distant star ever observed at a redshift of $z=1.49$ and with a magnification factor over 1000 \citep{Kelly2018}. The quest for highly magnified distant galaxies is of great relevance because it allows us to study the luminosity function (LF) of galaxies at high redshift, and learn about their formation history. Moreover, although LF from different groups are in good agreement on the global shape of the LF at M$_{UV} \lesssim -17$ mag, significant discrepancies have been found at the very faint-end, where detections are only feasible for highly-magnified sources. To illustrate this disagreement, on one hand, recent results suggest that the luminosity function presents a turnover at $z\approx 8-9$ \citep{Atek2015,Bouwens2015,Atek2018}. On the other hand, other authors did not find any indication of a turnover towards the faint-end of the luminosity function for $z \gtrsim 6$ \citep{Livermore2017,Ishigaki2018}. Future observations by the JWST will reveal fainter galaxies and probe the luminosity function up to $z \approx 12$. The findings made by JWST will constrain many evolutionary models, and perhaps rule out, or confirm, some exotic ones that predict that JWST will not find galaxies beyond $z=13$ \citep{Schive2014}. Observations of galaxies at $z>10$ will be relevant also to understand the reionization history of the Universe, providing independent constraints on the optical depth, critical, for instance, for future cosmological studies based on CMB data.  

The \textit{Hubble Frontier Fields} program \citep[hereafter HFF,][]{Lotz2017} represents the best effort to date to characterize with high precision some of the most significant lenses known so far. Six powerful lenses at redshift $0.3 \lesssim z \lesssim 0.55$ were observed to a depth of $m_{AB}\approx 29$ in the $r$-band. Each cluster contains of the order of one hundred multiply lensed images of distant background galaxies. Many of these multiply lensed images are unresolved, or very small, and lack spectroscopic information, making the identification of lensing systems a challenging task. On the other hand, a number of systems can be reliably identified thanks to morphological information, but more importantly, to spectroscopic information provided by HST's GRISM instrument \citep{Schmidt2014,Treu2015} and VLT's MUSE instrument \citep{Jauzac2016,Caminha2017} among others. The number of reliable systems per cluster with spectroscopic redshift ranges from approximately 10 to 35, resulting in a number of constraints (with secured redshift) above 50. In addition to these reliable systems, a second group of systems without spectroscopic redshifts, but with reliable photo-$z$ and gravitational lensing redshift, can be used to extend the number of systems per cluster and, consequently, to increase above 100 the number of constraints.

In this paper, we present lensing models for all six HFF clusters, and compute the probability of magnification for each one as a function of the magnification factor. Even though the number of available constraints is relatively large, different modeling techniques can result in models that are significantly different from each other. An intriguing effect is the fact that two lens models that have similar critical curves at a fixed redshift may differ by a factor 2 (or sometimes even more) in the predicted magnification at positions near the critical curves \citep{Bouwens2017,Diego2018a}. This discrepancy is due mostly to a different slope in the potential around the critical curves, resulting in models with shallower potentials producing larger magnifications near the critical curves. It is interesting to characterize this source of uncertainty, which should be taken into account when computing the expected number of lensed galaxies behind the clusters.  We address this issue by comparing our probability of lensing with the ones predicted by other models produced by different authors, but using similar strong lensing constraints. 

This work is organized as follows. 
Section~\ref{sec:models} introduces the free-form gravitational lensing models along with the observational data and two different methods to quantify the quality of the lensing models. 
Section~\ref{sec:maps} presents the probability of magnification. This section includes also a comparison with other public lens models of the same clusters, and examines its dependence with the source redshift. 
In section~\ref{sec:lensedgalaxies}, we examine how the lensing efficiency of the HFF clusters affects the observed number density of galaxies at high-$z$. 
Finally, section~\ref{sec:conclusions} summarizes our results and conclusions.

\section{Lensing reconstruction}
\label{sec:models}
In order to produce the lens models, we use the free-form code \wslap~\citep[see][for further details]{Diego2005,Diego2007,Sendra2014} which builds a lens model based on strong lensing data and, when available, weak lensing data. The mass distribution is built as a superposition of a diffuse component, plus a compact component that traces the light of the member galaxies (up to some truncation radius). The mass associated to the member galaxies in the compact component is assumed to be proportional to their luminosity following a given mass-to-light ($M/L$) ratio. In some cases, some member galaxies (like the Brightest Cluster Galaxy, BCG, or massive galaxies) can be considered in a separate layer (or compact component) allowing to have their own $M/L$ ratio, and adding additional free parameters to the lens models that correspond to the correction that needs to be applied to the fiducial $M/L$ ratio. The diffuse component, distributed as a superposition of Gaussian functions on a grid, is described by as many free parameters as grid points. The Gaussian functions can have equal widths (regular grid) or varying widths (multi-resolution grid). Multi-resolution grids are normally used when the cluster presents nearly spherical symmetry. This choice allows us to sample dense regions more heavily and to reduce drastically the number of cells needed to reproduce accurately the lensing properties of a cluster \citep{Diego2005}. For clusters with an irregular distribution, regular grids are the preferred option. The code optimizes the masses in the grid, and in the galaxies (i.e the compact component), by minimizing the square of a residual. This residual is  defined as the difference between the observed and predicted positions (and shear measurements when available).

\begin{table}
\begin{center}
\caption{Summary of the number of constraints in the the six HFF clusters. The second column indicates the field of view (\textit{fov}, in arcmin$^2$) considered to derive the lensing models. The third column shows the redshift of each cluster. Columns 4 and 5 correspond to the number of constraints (or systems, in parenthesis) included in the lensing models for the \textsc{gold} and the \textsc{all} sets, respectively. Note how for MACS 1149, while the number of systems is low (9), the number of constraints is still high because system 1 alone contains over 20 identifiable knots.}
\label{tb:numarcs}
\begin{tabular}{lcccc}
\hline
HFF &\textit{fov} & $z$ & \textsc{gold} & \textsc{all}\\
\hline
Abell 370 & $3.6\times3.6$ & 0.375 & 103 (36) & 122 (41)\\
Abell 2744 & $4.4\times4.4$ & 0.308 & 76 (24) & 180 (52)\\
Abells 1063 & $4.4\times4.4$  & 0.348 & 48 (18) & 91 (31)\\
MACS 0416 & $3.6\times3.6$ & 0.396 & 113 (35) & 158 (54)\\
MACS 0717 & $3.6\times3.6$ & 0.548 & 54 (9) & 72 (17)\\
MACS 1149 & $3.6\times3.6$ & 0.544 &  132 (9) & 154 (17)\\
\end{tabular}
\end{center}
\end{table}

The models presented in this study are derived using strong lensing data only. For each HFF cluster, we derive different sets of solutions depending on the multiple images considered to derive the mass models. Multiple images are identified in HST color images by seven independent teams (including ours, \textit{Frontier Fields Lensing Models v.4}). Different teams ranked the multiple images based on the availability of a spectroscopic redshift, and consistency between the multiple images (of each system) in terms of color, surface brightness, and morphology. Multiple images with available spectroscopic redshifts, and secure system associations are ranked as \textsc{gold}. Systems without spectroscopic  redshift (but reliable photometric redshift) but reliable identification based on color, morphology and consistency with a base lens model are ranked as \textsc{silver}. Finally, candidates ranked as \textsc{bronze} are usually unresolved systems without spectroscopic information for which neither secure redshift, nor morphology can be used to confirm the system but that show high consistency in the colors and relative positions (as predicted by preliminary models).

The first set of solutions is obtained with the \textsc{gold} candidates, while the second set of solutions is obtained after using all systems: \textsc{all=gold+silver+bronze}. The models are obtained after a large number of iterations (300K) in the optimization procedure. Additionally, for each cluster and data set (\textsc{gold} or \textsc{all}) we derive also a set of 100 solutions obtained after iterating 100K in the optimization per solution. This number of iterations (100K) is typically sufficient to achieve convergence and obtain reliable solutions \citep{Sendra2014}. For each of the 100 solutions, we vary the initial starting point in the optimization and the redshift of the systems with photometric redshifts (i.e., not \textsc{gold} candidates). The redshifts are sampled for each realization following a normal distribution with mean equal to the redshift of the system ($z_i$) and standard deviation equal to $0.2(1+z_i)$. The models presented in this work are publicly available through the MAST archive\footnote{https://archive.stsci.edu/prepds/frontier/lensmodels/}. Table \ref{tb:numarcs} shows the number of multiple images that are included in the lens models for each cluster. Appendix~\ref{app} contains the catalogs with the details on the multiple-image systems (\textsc{gold}, \textsc{silver} and \textsc{bronze}) as defined by the \textit{Frontier Fields Lensing Models v.4} collaboration. Hereafter, we will show results obtained with the \textsc{gold} candidates.

In the following subsections, we describe briefly each one of the six HFF clusters.

\subsection{Abell 370}
\label{sec:ABELL370}
This cluster presents two prominent BCG-type galaxies separated by roughly 190 kpc. This cluster is undergoing a merging process with significant extended X-ray emission in the region between the two BCGs, and local peaks near the BCG galaxies. 

The member galaxies are selected from \citet{Lagattuta2017} and \textsc{glass} \citep{Schmidt2014,Treu2015}. We included a total of 157 cluster members with $z = 0.375 \pm 0.02$ within the field of view of the HFF.

The mass distribution of the the lens consists on a superposition of three layers: one for the smallest member galaxies and two layers for each of the BCGs (North BCG and South BCG respectively).

The total number of constraints is equal to 122 candidates from 41 systems, from which 103 are \textsc{gold} candidates from 36 systems with spectroscopic redshifts.

\subsection{Abell 2744}
\label{sec:ABELL2744}
As in Abell 370 galaxy cluster, Abell 2744 is in a collisional phase with clear signs (from the X-ray emission) of a recent major merger event. The cluster presents also several prominent galaxies near the central region. Two of them are interpreted as BCG galaxies, while the other two prominent galaxies are located on both sides of the bright star that is clearly visible towards the Northwest of the BCG pair.

Member galaxies are selected from AstroDeep \citep{Merlin2016,Castellano2016}, \textsc{glass} \citep{Treu2015} and \citet{Owers2011}. We included a total of 403 cluster members with $z = 0.308 \pm 0.05$ within the field of view of the HFF.

We consider a total of five layers to describe mass distribution of the lens: one for the less bright galaxies, two layers for each of the BCG (center BCG and South BCG) and two layers for each one of the other two prominent galaxies located on the Northwest.

Arcs that present multiple knots are used with the knot information added as extra constraints. In particular, system 1 is described by 14 knots in 5 subsystems, while system 2 contains 20 knots from 5 subsystems. The total number of constraints is equal to 180 candidates from 53 systems, from which 76 are \textsc{gold} candidates from 24 systems with spectroscopic redshifts.

\subsection{Abells 1063}
\label{sec:ABELLS1063}
This cluster is relatively relaxed, presenting only one dominant BCG galaxy. 
The X-ray emission is centered on the BCG and shows no signs of recent merging activity. 

Member galaxies are selected from \citealt{Karman2017}, \textsc{glass} \citep{Schmidt2014,Treu2015} and \textsc{clash-vlt} (Mercurio et al. in prep.). We included a total of 338 cluster members with $z = 0.348 \pm 0.0135$.

For the compact component, we consider a total of two layers: one for central BCG galaxy and a second one for the remaining galaxies.

As before, arcs that present multiple knots are used with the knots added as additional constraints. In particular, system 14 is described by 5 knots in 3 subsystems but only for the set of solutions with \textsc{all} candidates given that the identified knots are not spectroscopically confirmed. We considered a total of 91 candidates from 31 systems, from which 48 are \textsc{gold} candidates from 18 systems with spectroscopic redshifts.

\subsection{MACS 0416}
\label{sec:MACS0416}
This lens is another example of a collisional cluster although possibly in an early phase of the collision \citep[see for instance][]{Diego2015}. 
Member galaxies are selected from VIMOS CLASH-VLT campaign \citep{Balestra2016} and the VLT/MUSE spectroscopic study \citep{Caminha2017}. We included 175 cluster member in the redshift range $z = (0.35, 0.45)$ (i.e., $\pm 0.05$ around the mean redshift of the cluster, $z=0.396$) with quality flags equal or greater than 3 (i.e., 168 objects with $100\%$ reliability and 7 objects with $>90\%$ reliability). In addition to these, we included 74 cluster members identified in \citet{Zitrin2013} based on F814W-F475W color. Therefore, we considered 249 cluster members with $z = 0.396 \pm 0.05$.

For the compact component of the mass distribution we consider four layers: one for the less massive galaxies, two layers for each of the BCG (North BCG and South BCG) and one layer for a foreground galaxy at $z=0.112$ in the Southwest region of the cluster, on top of the prominent arc.

Regarding arcs with multiple knots (added as extra constraints), system 1 is described by 3 subsystems with 2 knots each (previously systems 1 and 2); system 2 is described by 3 subsystems with 2 knots each (previously systems 3 and 4); and system 10 is described by 3 subsystems with 4 knots each (previously system 17). In summary, we included 158 candidates from 54 background systems; 113 are \textsc{gold} candidates from 35 systems with spectroscopic redshifts. 

\subsection{MACS 0717}
\label{sec:MACS0717}
This is the most complex cluster among the six clusters (and arguably one of the most complex clusters known in terms of number of subgroups) with at least three, and possibly four (or even more) subgroups caught in the middle of a massive collision. The cluster is highly disturbed with X-rays showing no clear correlation with the galaxies in the cluster.  No obvious BCG can be appreciated in the cluster. 

We included more a total of 338 cluster members in the field of view of the HFF taken from \citet{Richard2014},which combines a color-color selection (f435W-606W vs 606W-814W), and two color-magnitude selections (606W-814W vs 814W; 435W-606W vs 814W), calibrated by spectroscopically confirmed cluster members.

The compact component is divided into two layers: one for all the member galaxies in the cluster and one layer for a foreground galaxy at $z=0.154$ in the Southwest region of the cluster.

Arcs that present multiple knots are used with the knot information added as extra constraints: system 1 consists of a total of 21 knots in 5 subsystems (summing up all the knots, not 21 knots per subsystem); system 3 consists of 6 knots belonging to 3 subsystems; and system 5 consists of 9 knots in 3 subsystems. In total, we considered 72 candidates belonging to 17 systems; 54 candidates from 9 systems are labeled as \textsc{gold} candidates (i.e., with spectroscopic redshifts available).

\subsection{MACS 1149}
\label{sec:MACS1149}
MACS 1149 is a semi-relaxed cluster but with signs of recent activity. In particular, the peak of the X-ray emission presents a significant offset with respect to the BCG in the SE-NW direction. 

We considered 203 cluster members within the field of view of the HFF taken from \citet{Richard2014}, which combines a color-color selection (f435W-606W vs 606W-814W), and two color-magnitude selections (606W-814W vs 814W; 435W-606W vs 814W), calibrated by spectroscopically confirmed cluster members. We completed our cluster member catalog with cluster galaxies spectroscopically confirmed by recent MUSE observations \citep{Jauzac2016}.

We consider a total of three layers to describe the lens: one for the less bright galaxies; one the central BCG; and one for the bright galaxy on the North edge of the cluster. The modeling of MACS 1149 is based on a multi-resolution grid with 576 grid points.

Arcs that present multiple knots are used with the knot information added as extra constraints. In particular, the systems that include knot information are system 1 (98 knots in 6 subsystems, including the position of the multiply lenses SN Refsdal) and system 3 (15 knots in 5 subsystems). The total number of constraints is equal to 154 candidates from 17 systems, from which 132 are \textsc{gold} candidates from 9 systems.

\subsection{Quality of the lens models}
\label{sec:goodness}

\begin{table}
\begin{center}
\caption{Goodness of fit for the gravitational lensing models. The first column indicates the cluster name. Second column shows the dataset of multiple images included in the different lensing models (\textsc{gold} and \textsc{all}). Third columns shows the root mean square (rms) of the distance (in arcsec) between the observed and model-predicted positions of the multiple images computed in the lens plane. Last two columns shows the Einstein radius ($\theta_E$, in arcsec) for $z_s = 3$ and $z_s = 9$.}
\label{tb:rms}
\begin{tabular}{ccccc}
\hline
HFF  & Model & rms & $\theta_E$ ('') & $\theta_E$ ('')\\
 & & ('') & ($z_s = 3$) & ($z_s = 9$)\\
\hline
Abell 370 &\textsc{gold} & 0.75 & 39.8 & 45.7\\
Abell 370 &\textsc{all} & 0.93 & 38.4 & 46.1\\
\hline
Abell 2744 &\textsc{gold} & 0.77 & 25.5 & 29.0\\
Abell 2744 &\textsc{all} & 0.94 & 26.0 & 30.0\\
\hline
Abells 1063 &\textsc{gold} & 0.46 & 32.8 & 36.4\\
Abells 1063 &\textsc{all} & 0.98 & 33.6 & 38.0\\
\hline
MACS 0416 &\textsc{gold} & 0.62 & 29.2 & 32.5\\
MACS 0416 &\textsc{all} & 0.54 & 29.0 & 32.4\\
\hline
MACS 0717 &\textsc{gold} & 0.29 & 58.4 & 65.6\\
MACS 0717 &\textsc{all} & 0.43 & 58.0 & 66.3\\
\hline
MACS 1149 &\textsc{gold} & 0.84 & 24.3 & 28.6\\
MACS 1149 &\textsc{all} & 0.64 & 26.1 & 34.7\\

\end{tabular}
\end{center}
\end{table}

Quantifying the quality of a lens model is not trivial, and in many cases, is not even possible. Traditionally, the root-mean-square (or rms) of the difference between the predicted and observed positions of the arclets has been used as the figure of merit to quantify the quality of a particular model. This is valid approach but only to the first order. Since, inevitably, erroneous assumptions are always made when modeling a lens (including, for instance, assumptions on the symmetry of halos, ellipticities, a minimum mass scale for substructures, number of susbtructures), a model that is able to reproduce the arclet positions with an rms very close to zero, can fail to correspond to the correct underlying mass distribution. Instead, that model will represent a biased view of the true underlying mass, with the bias being proportional to the error made in the original assumptions.  For the particular case of \wslap, this was demonstrated in dramatic fashion in \citet{Ponente2011}, where the model that was able to best reproduce the arcs positions, showed a spurious ring of dark matter, similar to the alleged ring of dark matter around the cluster CL0024+17 \citep{Jee2007}, and probably due to a similar systematic effect. Addittionally, \citet{Ponente2011} found that a small rms is not a guarantee that the model is unbiased. In fact, tests performed with mock data demonstrated that rms smaller than 0.2 arcsec are possible but at the expense of introducing spurious fluctuations in the mass distribution. A similar conclusion was reached by \citet{Sendra2014} using simulated lensing data. Hence, the rms does not provide a reliable figure of merit to compare models when the rms of these models are all below $\sim 1$ arcsec. Nevertheless, as mentioned earlier, a first order estimate of the quality of the models can be provided by the rms, since an rms much larger than 1 arcsec may be indicative of significant biases in the model. In this section we report the rms between the observed ($\theta^{\rm{obs}}$) and model-predicted ($\theta^{\rm{mod}}$) positions of the multiple images in the lens plane, which is computed as follows:
\beq
rms = \sqrt{\frac{1}{N} \sum_{i=1}^{N} \left( \theta_i^{\rm{obs}} - \theta_i^{\rm{mod}} \right)^2},
\eeq
where the total number of multiple images is denoted by $N$. The values of the \textit{rms} for the different lensing models of the six HFF clusters are shown in table \ref{tb:rms}. As can be see from the table, all models have rms$\sim1$ arcsec. In most cases, the rms of the models with all arcs is larger pointing towards possible biases in some of the assumed redshifts of the arcs. A notable exception is MACS 1149 where the rms of the model with all the arcs is $\approx25\%$ smaller than the rms for the model with the gold ranked arcs. This may be because the model with the gold ranked arcs contained only seven systems while this number is increased to 17 when all arcs are included as constraints, increasing the robustness of the lens model. However, as mentioned above, having a smaller rms (once it is smaller than $\sim1$ arcsec) is no guarantee for a model being a better one.  

\begin{figure}
\centering
 \includegraphics[width=\columnwidth]{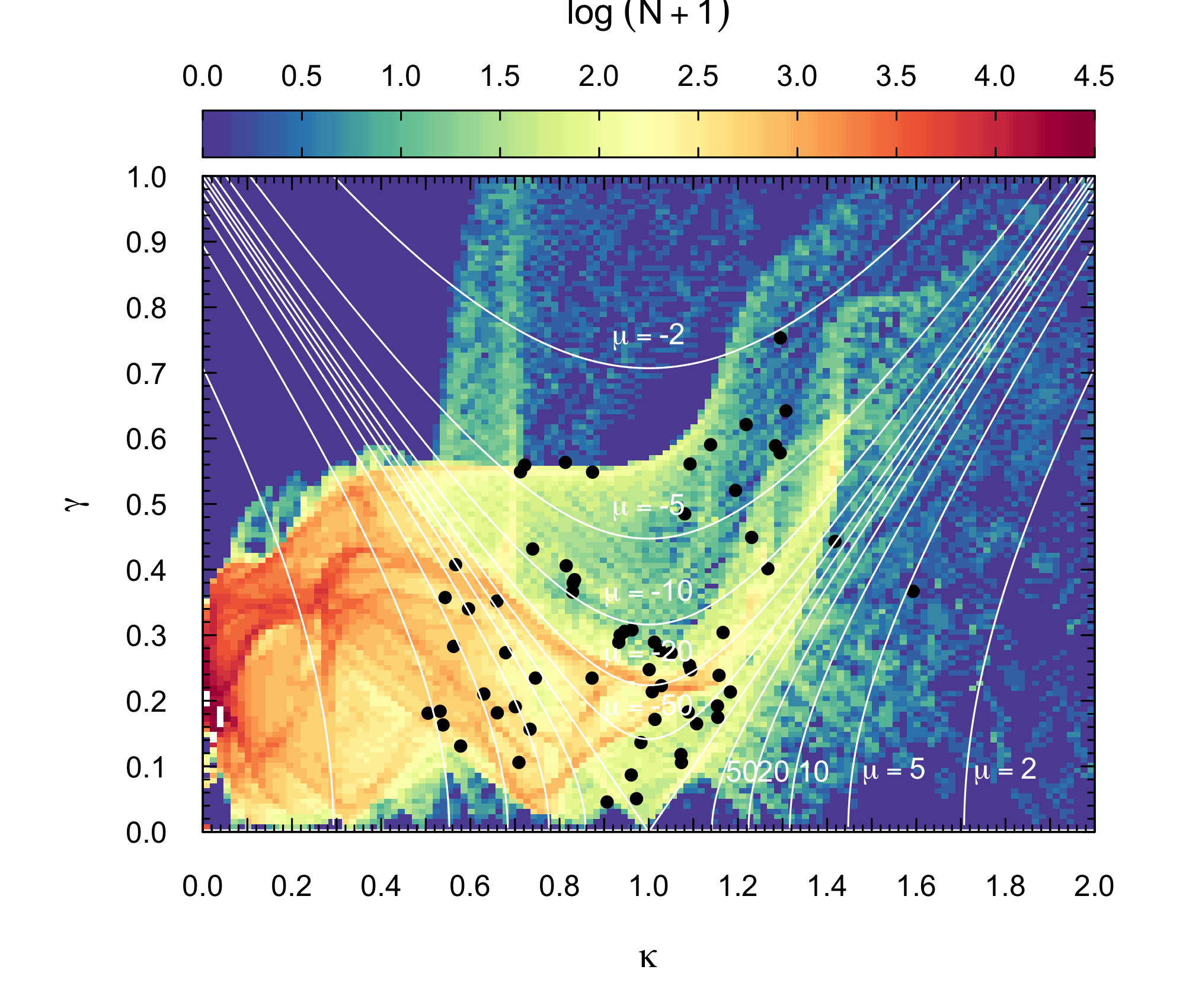}
 \caption{Lens' ghost for Abell 2744 (\textsc{gold} model) scaled to $D_{ls}/D_s = 1.0$. Color coding correspond to the log of the number of pixels (N) with a given convergence ($\kappa$) and shear ($\gamma$). Data points show values of convergence and shear for each of the multiple images included in the derivation of the lens model. Solid white lines indicate iso-contours of constant magnification: $| \mu | = (2,5,10,20,50)$. Note the caustic (or filamentary) structure and the relative good correspondence between the observed arcs (black dots) and the caustics. The dot at the rightmost position (around  $\kappa\approx 1.6$, and $\gamma\approx 0.35$) indicates potential issues with the model at the position of this arc (i.e. near the central region given the large value of  $\kappa$, and where the lensing constraints are more scarce).
\label{fig:lensghost}}
\end{figure}

\subsection{Lens' Ghosts. A self-consistency check}
\label{sec:check}

In this section we discuss a new, and alternative method that can be used to check the consistency of a given lens model. This method is based on what we define as the {\it lens' ghost} or simply the {\it ghost}. It follows from the principle that strongly lensed arcs are equally likely to appear anywhere within the region where multiple images form (assuming the observations are deep enough and neglecting projection effects with member galaxies that may hide small counter-images). This is a consequence of the conservation of surface brightness. It is important to note that all the counter-images of the same source should be detected, which is not true in all the cases. However, in practical terms, the counter-images that are more likely to be undetected are those that have low magnification (or are demagnified). If this is the case, any observed realization of strongly lensed arcs around a galaxy cluster should obey this uniformity law. By mapping the frequency of the values of the convergence, $\kappa$, and shear, $\gamma$, of a given lens model, the location of the arcs in the same space ($\kappa-\gamma$) should correlate. Departures from the correspondence between the frequency of the pairs $(\kappa,\gamma)$ and the locations of the observed arcs in the space of $\kappa-\gamma$ would reveal tensions between the model and the observations and, in particular, tensions with the (difficult to measure) magnification derived from the values of $\kappa$ and $\gamma$.

An example is shown in Figure~\ref{fig:lensghost}, for the particular case of Abell 2744. The ghost is computed at $D_{ls}/D_s = 1.0$ (i.e., independent of the redshift of the source $z_s$) and the values of $\kappa$ and $\gamma$, that are computed at the arc positions, are re-scaled from their original values at the corresponding redshift of the source to  $D_{ls}/D_s = 1.0$. Each lens model has a different ghost that can be considered as a fingerprint of the model (the real ghost of the cluster is unknown). The ghost shows caustics in the $\kappa-\gamma$ space which are associated with the distribution of substructures in the cluster. For instance, the ghost of a spherically symmetric lens would consist of a single curve. This follows from the perfect correspondence that exists between $\kappa$ and $\gamma$ (for circularly symmetric lenses $\gamma(r) = \bar{\kappa}(<r) - \kappa(r)$, where $\bar{\kappa}(<r)$ is the average convergence computed between $r=0$ and $r$). A model with no substructure, but with ellipticity, would have a ghost consisting of: 1) a curved band with high frequency that gets narrower towards the low values of $\kappa$ and $\gamma$ (that is, towards the bottom-left corner of Figure~\ref{fig:lensghost}); 2) two caustic-like regions (i.e, with a high frequency for the pairs of $\kappa$ and $\gamma$), one at the top of the band, and the other one at the bottom of the band. If substructure is added to the elliptical model, new and smaller caustics emerge inside the band, but also outside it. A reliable lens model should show consistency between the ghost and the distribution of arcs in the  $\kappa-\gamma$ space by tracing these caustics. Arcs that appear far away from the caustic regions in the  $\kappa-\gamma$ space may be indicative of systematics in the model in that part of the lens (i.e., where the arc is observed and with a corresponding value of  $\kappa$ and $\gamma$). This novel approach will be investigated in more detail in a future work including a full consistency check for the six HFF clusters. For the particular case of A2744 shown in Figure~\ref{fig:lensghost}, we appreciate how the position of the arcs (dots in the figure) tend to be located in regions of high frequency in the $\kappa-\gamma$ space, indicating that the lens model is in general consistent with the expected uniform distribution of the arcs locations in the lens plane. On the other hand, the arclet at $\kappa \approx 1.6$, $\gamma\approx0.35$ seems to fall relatively far from a caustic region indicating a possible systematic in this part of the lens (near the BCG galaxies).  

\section{Probability of magnification}
\label{sec:maps}
In this section we compute the probability of lensing for the six HFF clusters. This probability is defined in terms of the area in the source plane (in arcmin$^2$) that has a magnification larger than a given factor $\mu$ or between $\mu$ and $\mu + d\mu$. 

\subsection{Basics of Lensing}
The mapping between the image and source planes of a gravitational lens is described by the lens equation
\beq
\beta = \theta - \alpha \left(M, \theta \right),
\label{eq:lens}
\eeq
where $\beta$ is the (unobserved) position of the background source and $\theta$ is the observed position. The angle $\alpha \left(M, \theta \right)$ is the deflection produced by the mass distribution ($M$) acting as a gravitational lens at the position $\theta$. The lens equation may have more than one solution $\theta$ for a given position $\beta$. Therefore, the same source can be seen at several positions in the sky. The multiple occurrences of the same source are called counter-images. Additionally, gravitational lensing conserves the surface brightness, i.e., the surface brightness of a counter-image is identical to that of the source in the absence of the lens. The total flux of an observed image is proportional to its surface brightness and its subtended angle in the sky. Since lensing preserves the former quantity, the magnification factor $\mu$ is given by the ratio of the observed and original (i.e without lensing) areas, or similarly by the observed (i.e lensed) flux and the flux the source would have had without lensing:
\beq
\mu = \frac{d\Omega_{\theta}}{d\Omega_{\beta}},
\label{eq:mudef}
\eeq
where $d\Omega_{\theta}$ is the observed size of the counter-image and $d\Omega_{\beta}$ is the intrinsic size of the background source. The magnification can be also expressed as a function of the Jacobian between the unlensed and lensed coordinates, $\mathcal{A}$:
\beq
\mathcal{A} = \frac{\partial \beta}{\partial \theta} = \left(\begin{array}{cc}1 - \kappa - \gamma_1 & -\gamma_2 \\-\gamma_2 & 1 - \kappa - \gamma_1\end{array}\right),
\label{eq:jacobian}
\eeq
where $\kappa$ and $\gamma$ (convergence and shear respectively) are given by derivatives of the deflection field $\alpha \left(M, \theta \right)$. Then, the magnification is given by the inverse of the determinant of the Jacobian matrix: 
\beq
\mu = \frac{1}{\mathrm{det}~\mathcal{A}} = \frac{1}{\left( 1-\kappa \right)^2 - \gamma^2},
\label{eq:mu}
\eeq
where the magnitude of the shear is defined as $\gamma = \sqrt{\gamma_1^2 + \gamma_2^2}$. The magnification can be positive or negative depending on the parity of the counter-image. The parity of an image determines the specular orientation of the lensed images. The regions in the lens plane with different sign of Jacobian determinant are separated by the so-called critical curves (either tangential or radial). The Jacobian determinant vanishes at the critical curves where the magnification factor diverges. The critical curves map into caustics in the source plane through equation~\ref{eq:lens}.

The Taylor expansion of the lens equation \citep[see for instance][]{Diego2018a} around the critical curves yields (to second order):

\beq
\beta \simeq \frac{ \delta\theta^2}{2} \frac{d^2\beta}{d\theta^2},
\label{eq:expansion}
\eeq
where $\delta\theta=\theta-\theta_o$ is the distance to the critical curves in the lens plane (i.e, the critical curve is at $\theta=\theta_o$). Then the magnification near the critical curves can be expressed (to first order) as:

\beq
\mu = \frac{d\Omega_{\theta}}{d\Omega_{\beta}} \propto \delta\theta^{-1}. 
\label{eq:muarea}
\eeq
In the source plane, one gets,  
\beq
\mu \propto \sqrt{\delta\beta}\,^{-1},
\label{eq:muarea2}
\eeq
where $\delta\beta$ is the angular distance from the source to the caustic (in the source plane). 
The area (in the source plane) where a point-like source can be magnified by a factor larger than $\mu$ is defined as $\sigma_s(>\mu)$. Close to a caustic, they can be approximated by straight lines, and the probability of having magnification larger than $\mu$ is proportional to the probability of being at a distance smaller than $\delta\beta$ to the caustic, or according to equation~\ref{eq:muarea2},
$\sigma_s(>\mu) \propto \mu^{-2}$, a well known result \citep[see for instance][]{Schneider1992}. The differential area scales then as $\sigma_s (\mu) \propto \mu^{-3}$. Hereafter, with the subscript $s$ we denote that the area is computed in the source plane, while with the subscript $l$ we refer to the area in the lens (or image) plane. The above scalings describe regions that are close to a smooth caustic, that is, they are valid for large magnification factors ($\mu\approx10$ or above). The proportionality constants are related to the slope of the potential at the corresponding position of the critical curve,  which in turn is related to the Einstein radius. Shallower potentials (like those in clusters) result in larger normalization factors and lensing efficiencies. In real lenses, the presence of substructure introduces deviations from the smooth behaviour.  As shown in \citet{Cerny2018}, the lensing efficiency is not only a simple function of the lens mass, but it depends also on the lens redshift, elongation, and concentration. If we consider a spherically symmetric gravitational lens with a circular Einstein radius of $\theta_E$, the area (or probability if we normalize by the total area) in the lens plane above which a given magnification $\mu$ can be found, is given by the area in a region of thickness $2\delta\theta$ around the Einstein radius:

\beq
\sigma_l(>\mu) =  \sigma_l(<\delta\theta) = \int_{0}^{2\pi}  \mathrm{d}\phi \int_{\theta_E - \delta\theta}^{\theta_E + \delta\theta} \theta_E~\mathrm{d}\theta_E = 4\pi \theta_E \delta\theta.
\label{eq:Al_int}
\eeq
Since $\mu$ and $\delta\theta$ are related through equation~\ref{eq:muarea}, and the proportionality constant is related to $\theta_E$ \citep{Schneider1992}, we get:
\beq
\sigma_l(>\mu) \propto \theta_E^2 \mu^{-1} \propto \mu^{-1},
\label{eq:Al_theta}
\eeq
Since an area, $A$, in the image plane with magnification $\mu$, maps into an area $A/\mu$ in the source plane (equation~\ref{eq:mudef}), the corresponding area (or probability) in the source plane results in:

\beq
\sigma_s(>\mu) =  \sigma_l(>\mu) / \mu \propto \theta_E^2 \mu^{-2} \propto \mu^{-2}.
\label{eq:As_cum}
\eeq
The last expression, although derived for simple symmetric lenses with no substructure, and ignoring the contribution from the radial critical curve, is interesting because it shows how the lensing probability is expected to scale (to first order) as the square of the Einstein radius, and with the inverse of the magnification squared. We come back to this relation later on in this paper. In table \ref{tb:rms}, we also show the size of the \textit{effective} Einstein radius ($\theta_E$), as proposed by \citet{Redlich2012}, for a source at $z_s = 3$ and $z_s = 9$. If $A$ denotes the area enclosed by the tangential critical curve (and so in the lens plane), then the \textit{effective} Einstein radius is defined as the radius of the circle having the area $A$ (i.e., $\theta^2_E = A/ \pi$).

\subsection{Magnification in the source plane}
\label{sec:sourceplane}

\begin{figure*}
\centering
  \includegraphics[width=12cm]{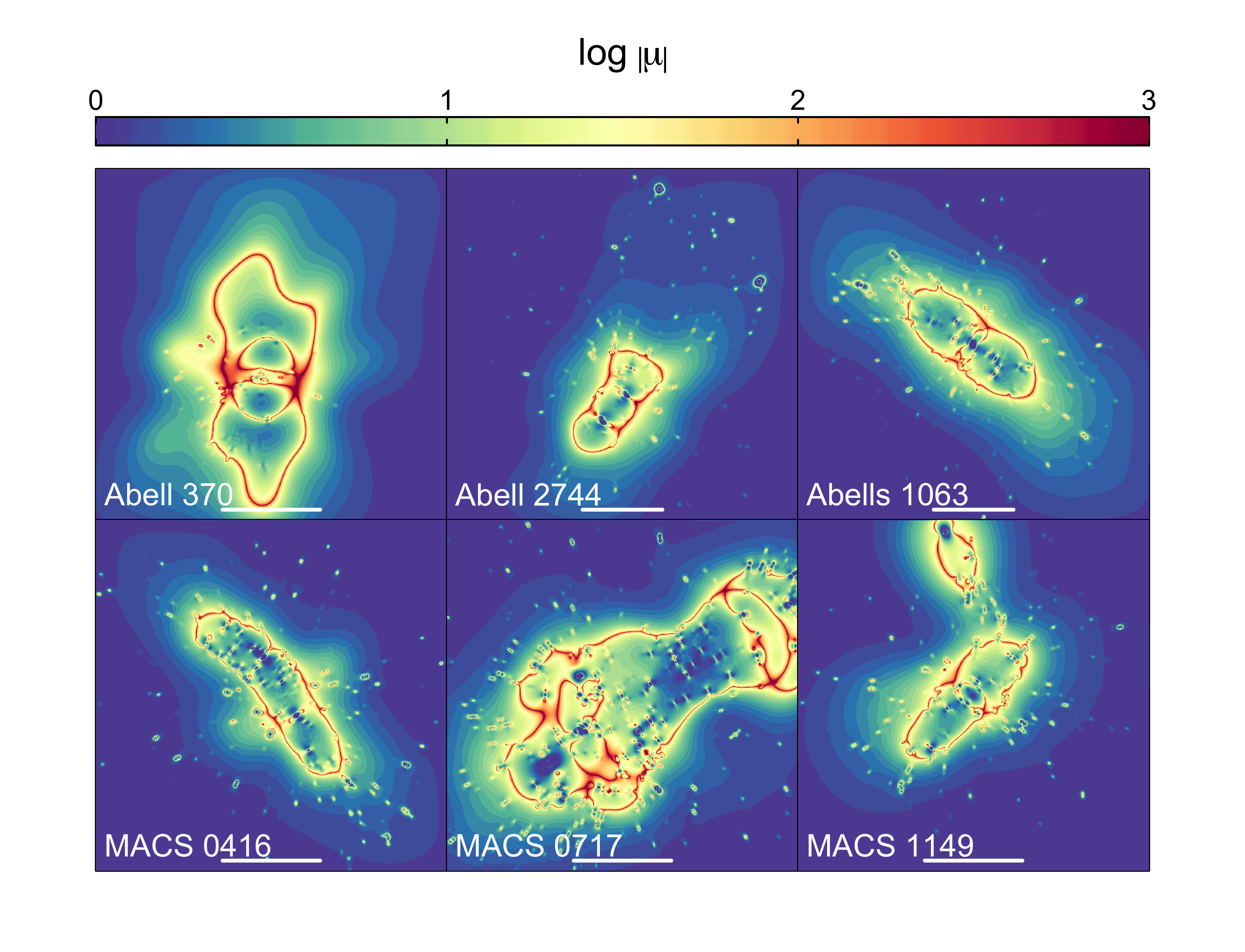}
\includegraphics[width=12cm]{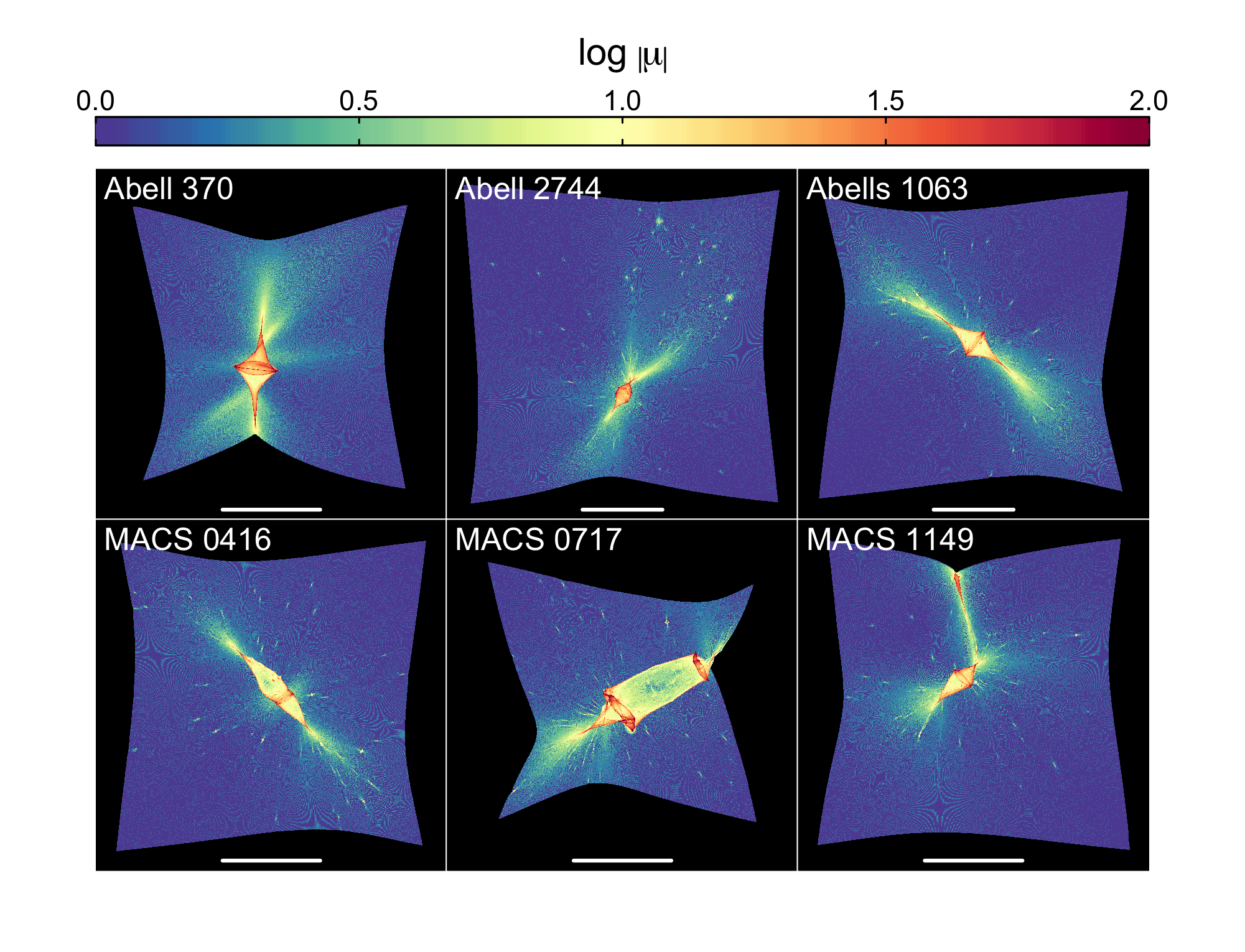}
 \caption{Magnification maps in the image plane (upper panels) and corresponding maps showing the caustics (bottom panel) obtained with \wslap~for the \textsc{gold} models of the six HFF clusters. The caustic maps are obtained using the \textit{source plane} method. The absolute value of the magnifications are expressed in log units up to $\rm{log}|\mu| = 3$ (upper panel) and $\rm{log}|\mu| = 2$ (bottom panels). The white lines correspond to 1 arcmin, while the total field of view (\textit{fov}) for each cluster is shown in table~\ref{tb:numarcs}. The source redshift is fixed at $z_s = 9.0$.
\label{fig:causmaps}}
\end{figure*}

Using the code \wslap~ we derive the lens model and from it the magnification maps in the lens plane for a given source redshift ($z_s$). To compare the different models at the same redshift of the source, we fix the source redshift at $z_s = 9.0$. In figure~\ref{fig:causmaps} we show the magnification maps for the entire field of view of the six HFF clusters, and for the \textsc{gold} models. These magnification maps have been computed at high resolution ($2048\times2048$ pix$^2$) by interpolating the original deflection angle maps ($512\times512$ pix$^2$) obtained with the \wslap~code. It is significantly much faster to interpolate the deflection field of $512\times512$ pix$^2$ to a resolution of $2048\times2048$ pix$^2$ than to recompute the deflection field with the \wslap~code at that resolution. The interpolation of the deflection field is expected to be reliable since the deflection field is smooth (except very close to the center of some galaxy members). To confirm our expectations we checked that the interpolation is valid for pixel sizes 16 ($4\times4$) times smaller than the original one, obtaining consistent values of the magnification after the interpolation. MACS 0717 is the cluster with the most complex structure and also the one with the largest tangential critical curve ($\theta_E \approx 66$''). We should note, that the supercritical region (at $z_s=9$) of MACS 0717 extends beyond the ACS field of view so the critical curve at this redshift must be even larger than the one shown in figure~\ref{fig:causmaps}.  The next cluster in terms of extension of the critical curve is Abell 370. The remaining four clusters have smaller critical curves, with extensions which are similar among the four of them. Note that A2744 and AS1063 have fields of view that are 22.2\% times larger than for the other clusters.  

\begin{figure}
\centering
 \includegraphics[width=\columnwidth]{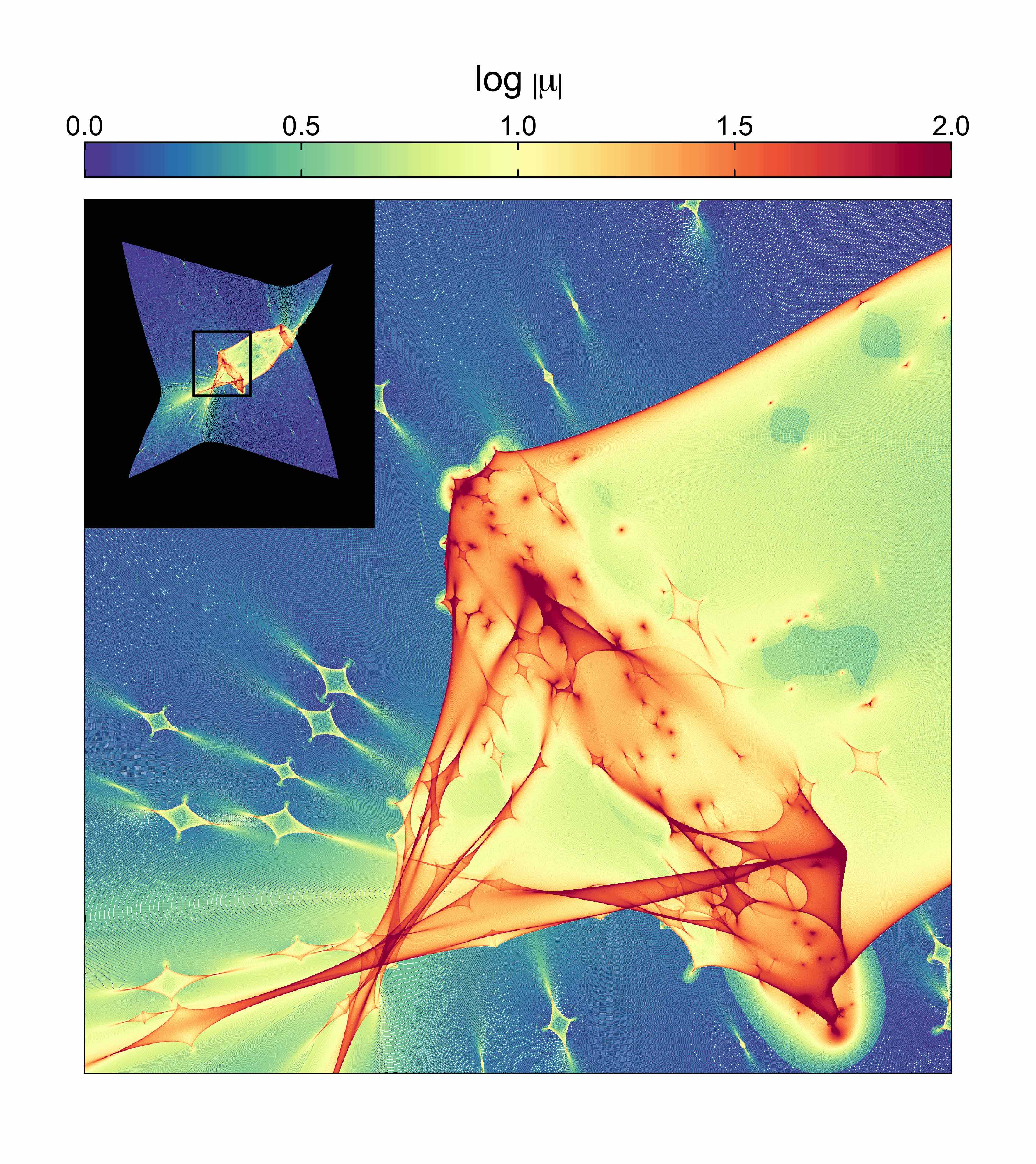}
 \caption{Zoom into a high magnification region in the HFF cluster MACS 0717. The enhanced region (black box in the top-left panel) corresponds to a patch of $\sim0.7\times0.7$ arcmin$^2$ extracted from a high-resolution map with $8192\times8192$ pix$^2$.
\label{fig:zoom_macs0717}}
\end{figure}

\begin{figure}
\centering
 \includegraphics[width=\columnwidth]{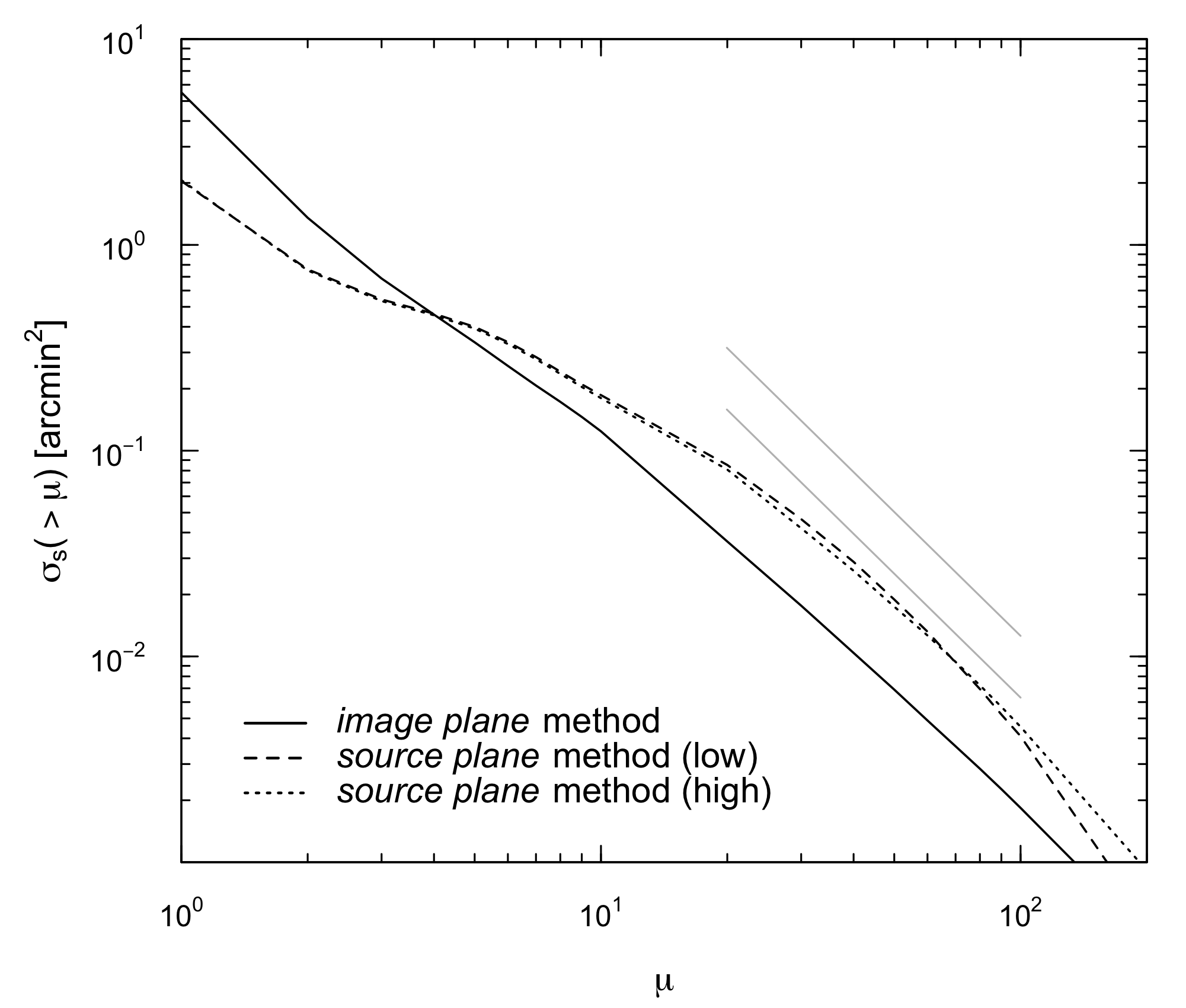}
 \caption{Area in the source plane covered by MACS 0717 for a magnification above a given threshold, $\sigma_s(>\mu)$. The source redshift is fixed at $z_s = 9.0$. Solid black line corresponds to $\sigma_s$ computed from $\sigma_l$ using the \textit{image plane} method (equation~\ref{eq:As_cum}). Dashed and dotted lines show the $\sigma_s$ computed with the \textit{source plane} method (equation~\ref{eq:mu_ray}) for different grid resolution (in pix$^2$): $2048\times2048$ (\textit{low}) and $8192\times8192$ (\textit{high}), respectively.  The gray solid lines indicate the dependence $\mu^{-2}$ (equation~\ref{eq:As_cum}) with a separation equal to 0.3 in log-scale, which corresponds to a factor $\approx 2$.
\label{fig:resolution_macs0717}}
\end{figure}

\begin{figure*}
\centering
 \includegraphics[width=13cm,angle=0.0]{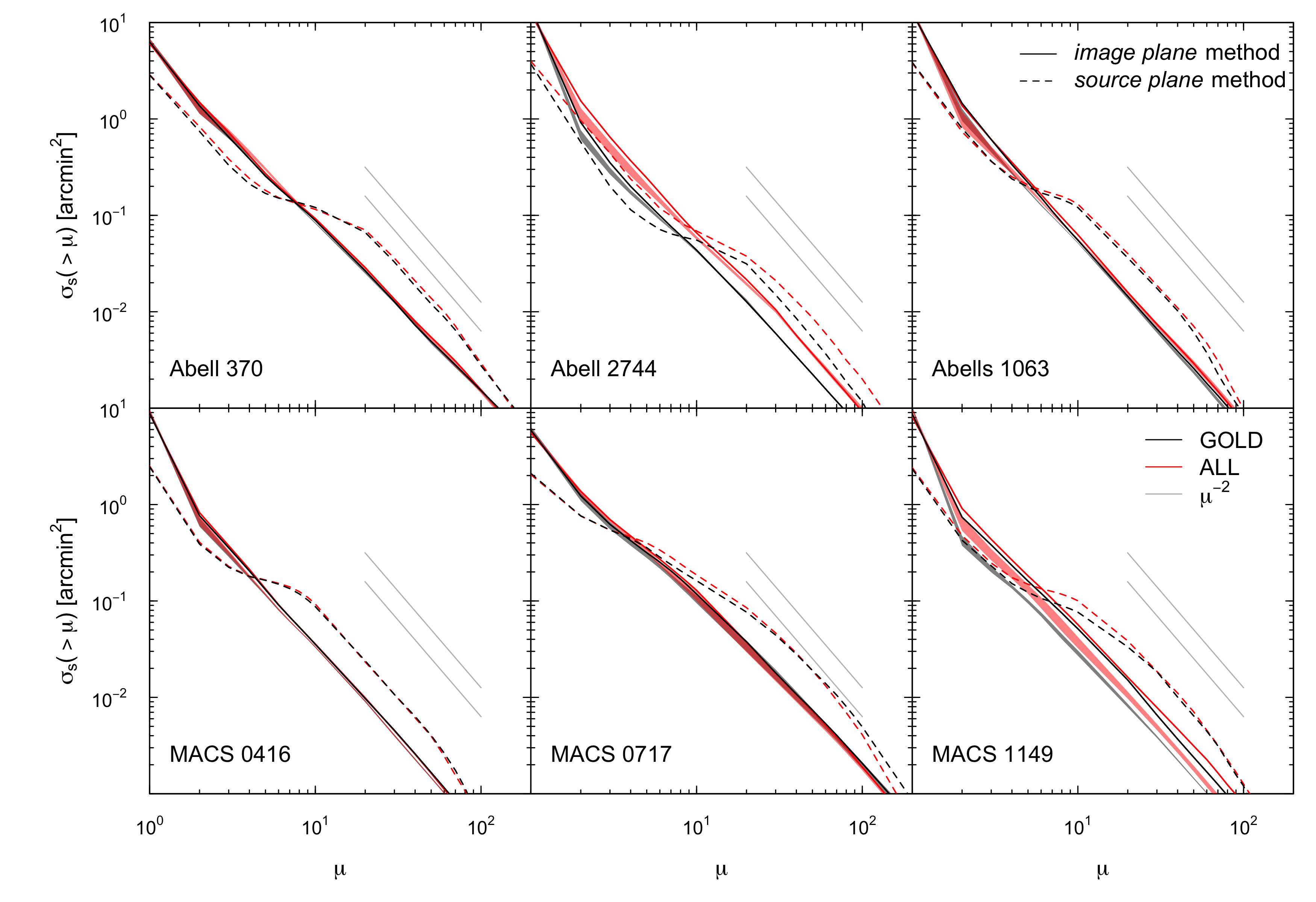}
 \caption{Area in the source plane with a magnification above a given threshold, $\sigma_s(>\mu)$, covered by the six HFF clusters. Solid lines correspond to the results obtained using the \textit{image plane} method, while dashed lines show the results for the \textit{source plane} method. Black lines indicate the results for the \textsc{gold} models, while red lines correspond to the \textsc{all} models with 300k iterations in the optimization procedure. Shaded regions in grey and red correspond to the 2-$\sigma$ values for the 100 solutions obtained after iterating 100K in the optimization per solution for the \textsc{gold} and the \textsc{all} models, respectively. The gray solid lines indicate the dependence $\mu^{-2}$ (equation~\ref{eq:As_cum}) with a separation equal to 0.3 in log-scale, which corresponds to a factor $\approx 2$. The source redshift is fixed at $z_s = 9.0$. All the data necessary to reproduce the solid and dashed curves of area above a given magnification in the source plane will be publicly available as online material.
 \label{fig:As_all}}
\end{figure*}

From the lens models we compute the area in the source plane that is magnified by a factor larger than $\mu$, also denoted as lensing efficiency \citep{Wong2012}. This is needed, for instance, to estimate the expected number of observed galaxies above a given detection threshold. If distant galaxies get magnified by a galaxy cluster, their flux will increase by a factor $\mu$, or similarly, they will gain $2.5\mathrm{log}_{10}~(\mu)$ apparent magnitudes. The number of galaxies above a given limiting magnitude can then be computed after integrating  the galaxy luminosity function convolved with the lensing efficiency (see section~\ref{sec:lensedgalaxies}).
   
\subsubsection{Methodology}
\label{sec:methods}
   
The area in the source plane with a given magnification can be computed in different ways, depending on how the magnification is interpreted. One may consider the magnification of a background source, that is, the ratio between the (observed) total flux in all counter-images and the (unobserved) flux of the unlensed source. This may be useful when one can not resolve the multiple images and only the total flux is observed. For instance, when two or more counter-images are magnified by very large factors and are too close to each other to be resolved or when the experiment lacks the angular resolution to resolve multiple images (such as low resolution data obtained with the Herschel Space Observatory). Alternatively, one may just consider the magnification of a single counter-image defined as the observed flux of that counter-image divided by the flux of the background source that would be obtained without lensing. Examples of this are situations where only the brightest counter-image is being observed. Depending on the situation, one may be more interested in the former or the later definition. For instance, if one is observing distant IR sources with low resolution experiments like the Planck telescope or the Herschel Space Observatory, one may be more interested in the total flux since the counter-images are unresolved. Another example can be found in the Icarus event, where multiple counter-images of a star at $z=1.49$ are formed by the combined lensing effect of a cluster plus microlenses \citep{Kelly2018,Diego2018a}. The counter-images are spread over a region not larger than a few milliarcseconds and, hence, they are all unresolved. Observations can only record changes in the total flux, which is given by the sum of all counter-images. On the contrary, if one is observing distant galaxies with a high-resolution camera, like WFC3 in HST, one may be more interested in the probability of just one of the counter-images being bright enough to be detected.

In this study, we compute the magnification in the source plane using two different techniques. We denote the \textit{source plane} method as computing the total magnification of the source plane, and denote the \textit{image plane} method as computing the magnification of individual counter-images. On one hand, for the \textit{source plane} method, pixel positions in the image plane are assigned to corresponding pixels in the source plane using the lens equation (equation~\ref{eq:lens}). In general, multiple pixels in the image plane land into the same pixel in the source plane. Therefore, the magnification in a pixel of the source plane is simply given by the number of pixels from the image plane that land in that pixel:

\begin{equation}
\mu_{i,j} = \sum_{u,v}(\delta_{l} / \delta{s}),
\label{eq:mu_ray}
\end{equation}
where $\mu_{i,j}$ denotes the magnification in the source plane at the pixel $(i,j)$, $\sum_{u,v}$ extends over the $(u,v)$ pixels from the lens plane that project (after de-lensing) into the pixel $(i,j)$ in the source plane, and $\delta_l$ and $\delta_s$ correspond to the pixel size in the lens and the source plane, respectively. We assume that pixels in the lens and source planes have the same size ($\delta_l \equiv \delta_s$). Therefore, when the value $\mu_{i,j}$ is mapped back to all the image plane pixels $(u,v)$ that image it, the $\mu$ values in the image plane derived with this \textit{source plane} method will then contain the total magnification (i.e, the sum of the magnification of all the counter-images of the same given background source). On the other hand, in the \textit{image plane} method,  each pixel in the lens plane is assigned magnification $\mu$ from equation~\ref{eq:mu} and is then assigned to an area in the source plane that is $\mu$ times smaller (following equation~\ref{eq:mudef}). This method is inexpensive and has been extensively used in literature \citep[see for instance][]{Johnson2014,Richard2014,Jauzac2015}, but does not recover the total magnification properly since it ignores the multiplicity of images.

\begin{table}
\begin{center}
\caption{Area in the source plane above a given magnification (in arcmin$^2$) covered by the six HFF clusters for $\mu>30$ and $z_s = 9$. First column indicates the cluster name. Second and third columns correspond to the \textit{image plane} and the \textit{source plane} methods, respectively. Fourth column shows the ratio of the second to the first column, denoted as $\sigma^s_s / \sigma^i_s$.}
\label{tb:As}
\begin{tabular}{cccc}

\hline
HFF & $\sigma_s(>\mu=30)$ & $\sigma_s(>\mu=30)$ & $\sigma^s_s / \sigma^i_s$\\
& \textit{image plane} & \textit{source plane}\\

\hline
Abell 370 & 0.0130 & 0.033 & 2.6\\
Abell 2744 & 0.0060 & 0.015 & 2.5\\
Abell S1063 & 0.0066 & 0.018 & 2.6\\
MACS 0416 & 0.0046 & 0.011 & 2.4\\
MACS 0717 & 0.0180 & 0.044 & 2.5\\
MACS 1149 & 0.0066 & 0.018 & 2.8\\
\end{tabular}
\end{center}
\end{table}

The relation between the two methods can be better understood at large magnification factors. If the magnification is large, the total magnification is usually dominated by the sum of magnifications of the two largest images, lying very close to a critical curve, and both with very similar magnifications. The \textit{image plane} method would assign to each image a magnification $\mu$, but it would count the area with this magnification twice (one for each counter-image).  This area would then be divided by the factor $\mu$ to compute the area (or probability) in the source plane. With the \textit{source plane} method, the area subtended by both counter-images  would overlap into the same area in the source plane. The pixels in the source plane are counted only once, as opposed to the \textit{image plane} method that would count them twice, but the magnification in that pixel would not be $\mu$, but $2\mu$ instead. In other words, if one takes the area above magnification $\mu$ computed using the \textit{image plane} method and divides that area by a factor 2, the resulting area would correspond to the area obtained using the \textit{source plane} method but computed at magnification $2\mu$. Consequently, if we plot the area in the source plane above a given magnification, $\sigma_s (>\mu)$, as a function of $\mu$ in log-log space, and assuming that $\sigma_s (>\mu) \propto \mu^{-2}$, the curves $\sigma_s (>\mu)$ for the \textit{source plane} method and the \textit{image plane} method should be separated by a factor of 2. This relation is demonstrated in \citet{Diego2018b}, where the two methods are compared using analytical lens models. In more realistic models, rich in substructure, this factor 2 is maintained only approximately as shown later. 

The bottom panel in figure~\ref{fig:causmaps} shows the magnification maps in the source plane for each HFF cluster using the \textit{source plane} method. The caustics are clearly visible as regions of extreme magnification. These maps are derived for a resolution of $2048\times2048$ pix$^2$. In order to illustrate the complexity of the caustics,  we show in figure~\ref{fig:zoom_macs0717} a zoom into a high magnification region of the cluster MACS 0717. The zoomed area covers a region of $\sim0.7\times0.7$ arcmin$^2$ extracted from a high-resolution map with $8192\times8192$ pix$^2$. The superposition of several caustics produced by the mass distribution, is clearly visible. Regions with overlapping caustics make great targets to study caustic crossing events \citep{Diego2018a,Diego2018b,Kelly2018,Windhorst2018} since the probability of a caustic crossing increases if the source plane is populated with bright stars moving towards the web of caustics. For magnifications larger than a few hundred, microlenses from the intracluster medium start to play a dominant role in the probability of magnification.  Microlenses near critical curves make it more likely to cross a microcaustic, momentarily boosting the magnifications up to several thousand. Future observations with JWST are expected to reach the required depth to see more caustic crossings of background luminous stars, including (hopefully) the first Pop III stars \citep{Windhorst2018}.

\begin{figure}
\centering
 \includegraphics[width=\columnwidth]{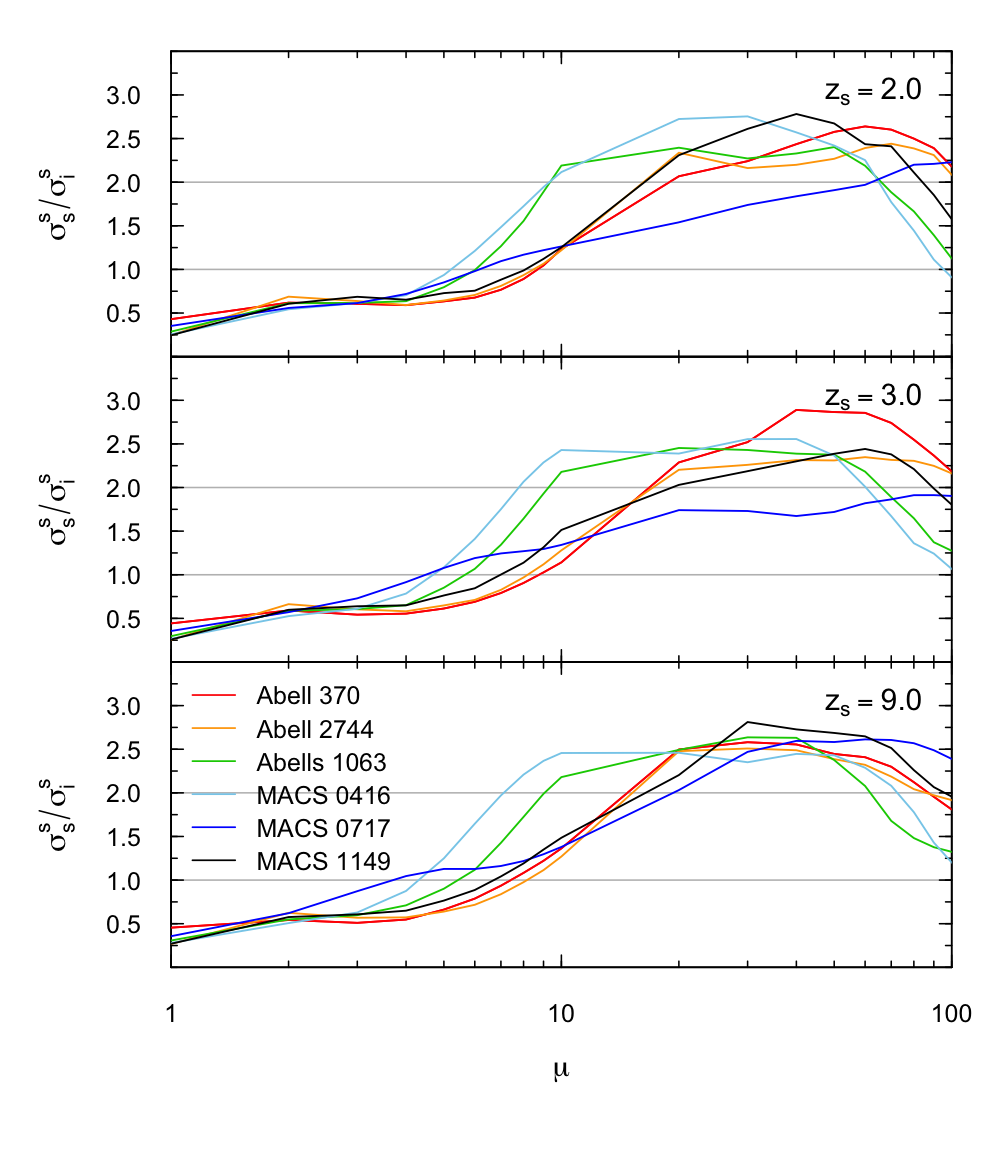}
 \caption{Ratio of the \textit{source plane} to the \textit{image plane} methods for $\sigma_s (>\mu)$, denoted as $\sigma^s_s / \sigma^i_s$ for $z_s =$ 2, 3 and 9. Only the ratios for the \textsc{gold} models are shown (i.e., ratio of the dashed black line and the solid black line in figure~\ref{fig:As_all}). Color lines correspond to each HFF cluster.
  \label{fig:ratio_As_all}}
\end{figure}

Computing the magnification maps in the source plane with the \textit{source plane} method has one drawback. Square pixels in the image plane map into irregularly shaped pixels in the source plane. At large magnifications, the pixels in the source plane become very small and thin, and one needs to rely on numerous interpolations to fill the source plane and create a uniform map without holes. These interpolations become increasingly more numerous as the magnification increases resulting in a very expensive computational process. A faster solution is to consider pixels in the source plane of similar size to those in the image plane so the number of interpolations needed is minimal. This, however, limits the maximum magnification that can be computed with accuracy, since at large magnifications pixels in the image plane (with large but significantly different magnification factors), map into the same pixel in the source plane. In these situations, and below the scale of the pixel, some information is lost and the lensing probability can not be computed with precision. This loss of information effect results in a bias in the area of the source plane at large magnification factors. 
In figure~\ref{fig:resolution_macs0717}, we show an example of the loss of information effect for MACS 0717 (only for the \textsc{all} mass models) where we show the $\sigma_s (>\mu)$ for the \textit{image plane} and the \textit{source plane} methods. The black solid line corresponds to the area derived with the \textit{image plane} method, where the area at a given $\mu \pm \delta\mu$ is computed in the image plane and divided by $\mu$ (following equation~\ref{eq:As_cum}) to account for the equivalent area in the source plane. The dashed, dotted and dashed-dotted lines show the area computed using the \textit{source plane} method for two different grid resolutions. Note how by increasing the resolution in the source plane, the expected $\mu^{-2}$ behavior is maintained for $\mu \gtrsim 100$.

\begin{figure*}
\centering
 \includegraphics[width=13cm,angle=0.0]{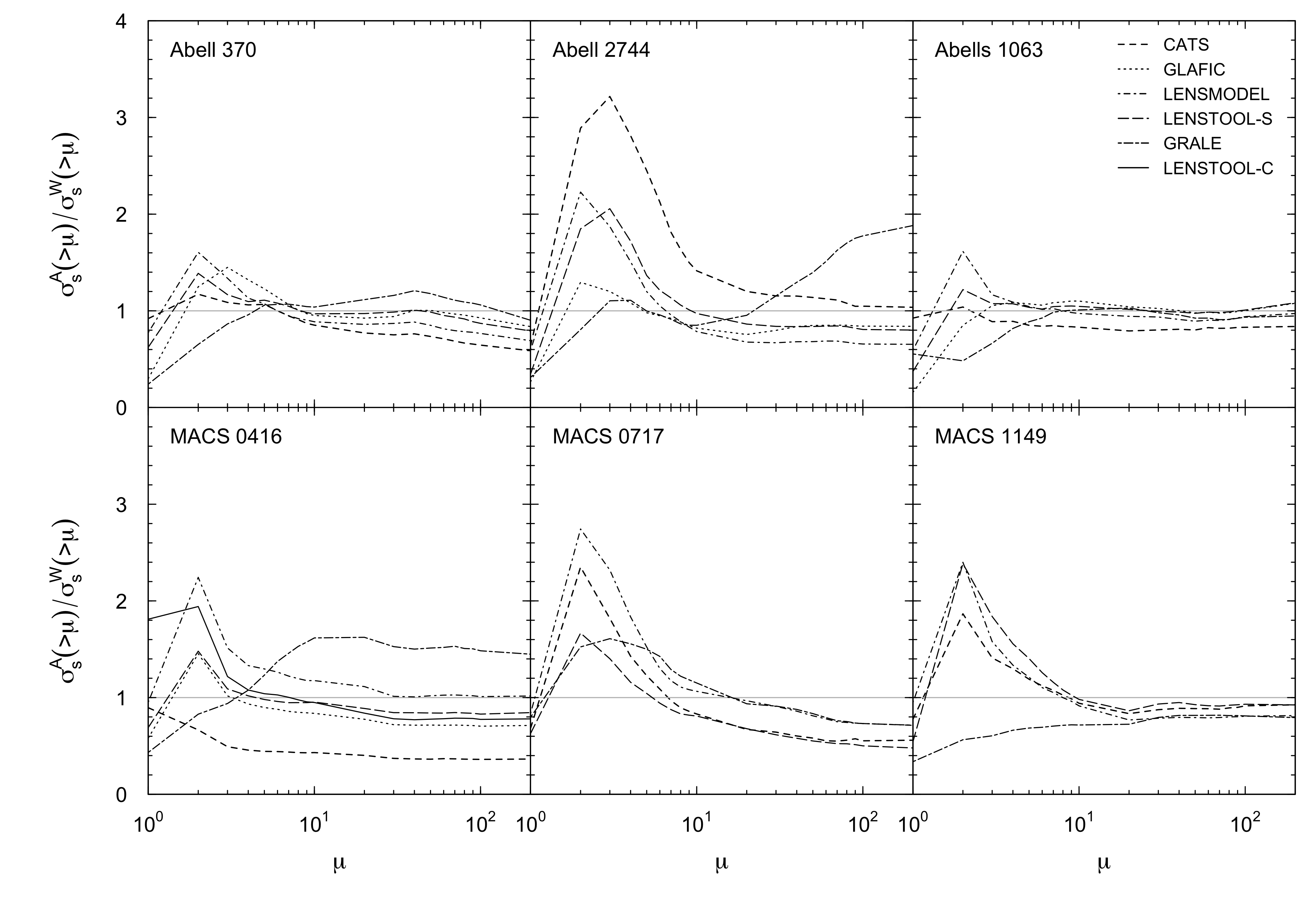}
 \caption{Ratio of $\sigma_s(>\mu)$ for the different lens model teams in the \textit{Frontier Fields Lensing Models v.4} collaboration, denoted as $\sigma_s(>\mu)^{A}$, with respect to the results of \wslap~model, denoted as $\sigma_s(>\mu)^{W}$. Results are shown only for the \textit{image plane} method and for a source redshift fixed at $z_s = 9.0$.
 \label{fig:As_teams_ratio}}
\end{figure*}

\subsubsection{Results}
\label{sec:results}

In figure~\ref{fig:As_all}, we show $\sigma_s(>\mu)$ for the all the derived mass models (\textsc{gold} and \textsc{all} models for 300k and 100k iterations) of the six HFF clusters considering a source at $z_s = 9$. Solid lines denote the results derived using the \textit{image plane} method, while dashed lines correspond to the results obtained by the \textit{source plane} method. Although some differences are expected to be found between the predictions for the \textsc{gold} and the \textsc{all} models, the results here presented are consistent for all the HFF clusters with the exception of Abell 2744. The predictions for the \textsc{all} models in Abell 2744 introduces some extra mass around the two prominent galaxies in the Northwest to predict the positions of the multiple-image systems 15 and 16 around them. Besides, systems 14, 36 and 38 also lead to an extension of the critical line towards the North, including a few areas of large magnification. In the case of MACS 1149, the differences in $\sigma_s(>\mu)$ are more evident when comparing the model predictions for the 300k iterations and the corresponding shades regions for the 100k iterations. This is mainly due to the mass assigned in the models with 300k iterations to the layer that contains the bright galaxy on the North edge of the cluster in order to reproduce the configuration of system 6 around it. Besides, given that this cluster was modeled using the multi-resolution grid, the North clump falls in a large pixel size (compared to pixels closer to the cluster center) for which the minimization procedure seems to not have properly converged after 100k iterations. All the data necessary to reproduce the curves of area above a given magnification in the source plane will be publicly available as online material.

The ratio of the \textit{source plane} to the \textit{image plane} methods for $\sigma_s (>\mu)$, denoted as $\sigma^s_s / \sigma^i_s$, are shown in figure~\ref{fig:ratio_As_all} for $z_s = (2.0, 3.0, 9.0)$. These ratios are below $\sim 3.0$ for any given $\mu$ and $z_s$. The connection between the surface in the source plane obtained with both methods was briefly described earlier. For high magnification (i.e., $\mu \gtrsim 20$), where both curves follow the $\mu^{-2}$ relation, the \textit{image plane} method is counting at least twice the area of the \textit{source plane} method, but it only accounts for the magnification of one of the counter-images. At these high magnifications, the majority of the flux of a given source is divided into two counter-images, each one carrying roughly half of the total magnification. Consequently, the ratio $\sigma^s_s / \sigma^i_s$ at $\mu \gtrsim 20$ is expected to be approximately a factor of 2, as we previously mentioned. Nevertheless, the number of counter-images that form from a given source is an odd number, so the additional counter-images will contribute to the total magnification in the \textit{source plane} method, but with a a much lower magnification than the two most magnified counter-images. This translates into a ratio $\sigma^s_s / \sigma^i_s$ slightly larger than 2. At low magnification ($\mu \lesssim 5$), the ratio $\sigma^s_s / \sigma^i_s$ tend to be lower than 1 for all the clusters and source redshifts analyzed. This is again due to the different definition of the magnification for the \textit{image plane} and \textit{source plane} methods:
the \textit{image plane} method yields a larger total source plane area because it counts multiply-imaged regions multiple times. In table~\ref{tb:As}, we show the values of $\sigma_s(>\mu=30)$ for the \textit{source plane} and the \textit{image plane} methods along with the ratio $\sigma^s_s / \sigma^i_s$ at $\mu > 30$ for a source at $z_s = 9$.

\subsubsection{Comparison with other model predictions}
\label{sec:team_comparison}

To check the consistency of our lensing model with other alternative models, we present a comparison in terms of $\sigma_s(>\mu)$ with the rest of the teams involved in the \textit{Frontier Fields Lensing Models v.4} collaboration, for which the lens models are publicly available at MAST archive\footnote{https://archive.stsci.edu/prepds/frontier/lensmodels/}. In figure~\ref{fig:As_teams_ratio}, we show the ratio of lensing efficiencies between alternative models and ours, 
$\sigma_s(>\mu)^{A}/\sigma_s(>\mu)^{W}$, where $\sigma_s(>\mu)^{A}$ is the lens efficiency from the alternative model and $\sigma_s(>\mu)^{W}$ is the corresponding efficiency of the \wslap~model.
The best agreement between models is found at large magnifications, specially for the two clusters that are the most relaxed ones, AS1063 and MACS 1149. The cases showing the largest dispersion between teams are Abell 2744, MACS 0416 and MACS 0717. At low magnifications ($\mu \lesssim 3$), the differences are typically the largest and up to a factor 3-4. This is expected when comparing our model with those derived from parametric methods since the \wslap~does not constrain the mass beyond the region covered by the lensing constraints (i.e, in the region where relatively small magnification factors are found). Parametric methods, instead, assume a fiducial profile in this regime, usually a Navarro-Frenk-White (NFW) profile, which typically results in larger magnifications beyond the constrained region.  
For intermediate and high magnifications, the agreement between different teams is better but still differences of a factor 2 can be 
seen. This is a well known effect where models that predict almost identical critical curves can differ by a factor 2 in the predicted magnification at small distances from the critical curve \citep[see for instance the discussion in section 8.1 of][]{Diego2018a}. 

The case showing the larger scatter between teams at high magnification is MACS 0416. For the rest of the clusters, \wslap~model falls in between the predictions for the rest of the teams, with the exception of MACS 0717 where \wslap~predictions for $\sigma_s(>\mu)$ are about 20\% larger than for the rest of the teams. The tendency for \wslap~to predict larger magnification factors near critical curves is a consequence of the intrinsically smooth potentials that emerge from the superposition of the Gaussians at the grid points in the lens plane. 

We also compare our lens models with those derived before the HFF data was made available. We derived the lensing efficiency of the HFF clusters by interpolating the curves of $\sigma_\mu (> \mu)$ at $\mu = 30$ (using the \textit{image plane} method) for the lensing models presented in \citet{Richard2014}, and compare them with our $\sigma_s(\mu > 30)$ for the \textit{image plane} method given in table~\ref{tb:As}. For all the cases with the exception of Abell 2744, the values of $\sigma_s(>\mu=30)$ have increased by a factor of 1.4 for MACS 1149 to 2.3 for Abell 370 and Abell 1063. Although the supercritical region of MACS 0717 is not fully included in the ACS field of view ($3.6 \times 3.6$ arcmin$^2$), it is the most efficient gravitational lens of the six HFF clusters followed by Abell 370 and Abells 1063. The same is observed when looking at the size of the effective Einstein radius,  which for MACS 0717 is $\theta_E \approx 66$ arcsec (see table~\ref{tb:rms}), that is, the largest of the HFF clusters.  It should also be noted that both Abell 370 and Abells 1063 are at a lower redshift than MACS 0717. 

\subsubsection{Dependence with source redshift}
\label{sec:source_dependence}

\begin{figure}
\centering
 \includegraphics[width=\columnwidth]{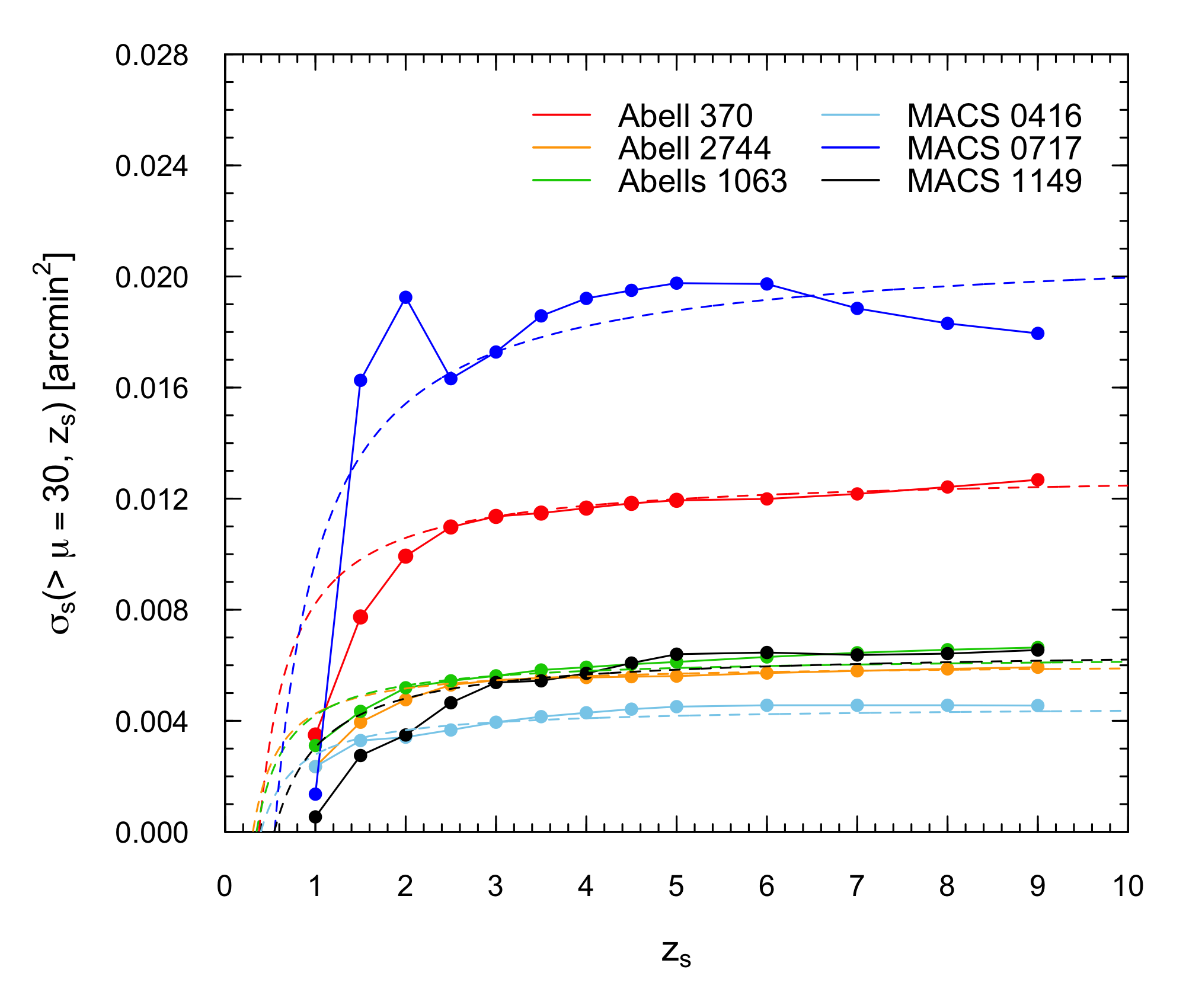}
 \caption{Normalization parameter $\sigma_0(z_s)$ as a function of the source redshift ($z_s$) for the six HFF clusters. Solid lines and circles correspond to the computed $\sigma_0(z_s)$ for the six HFF clusters. Dashed lines show the relation $D_{ls}/(D_l~D_s)$ (inverse of the \textit{effective lensing distance}) normalized by $\sigma_s(>\mu=30, z_s = 3)$.
  \label{fig:A0_source}}
\end{figure}

\begin{figure}
\centering
 \includegraphics[width=\columnwidth]{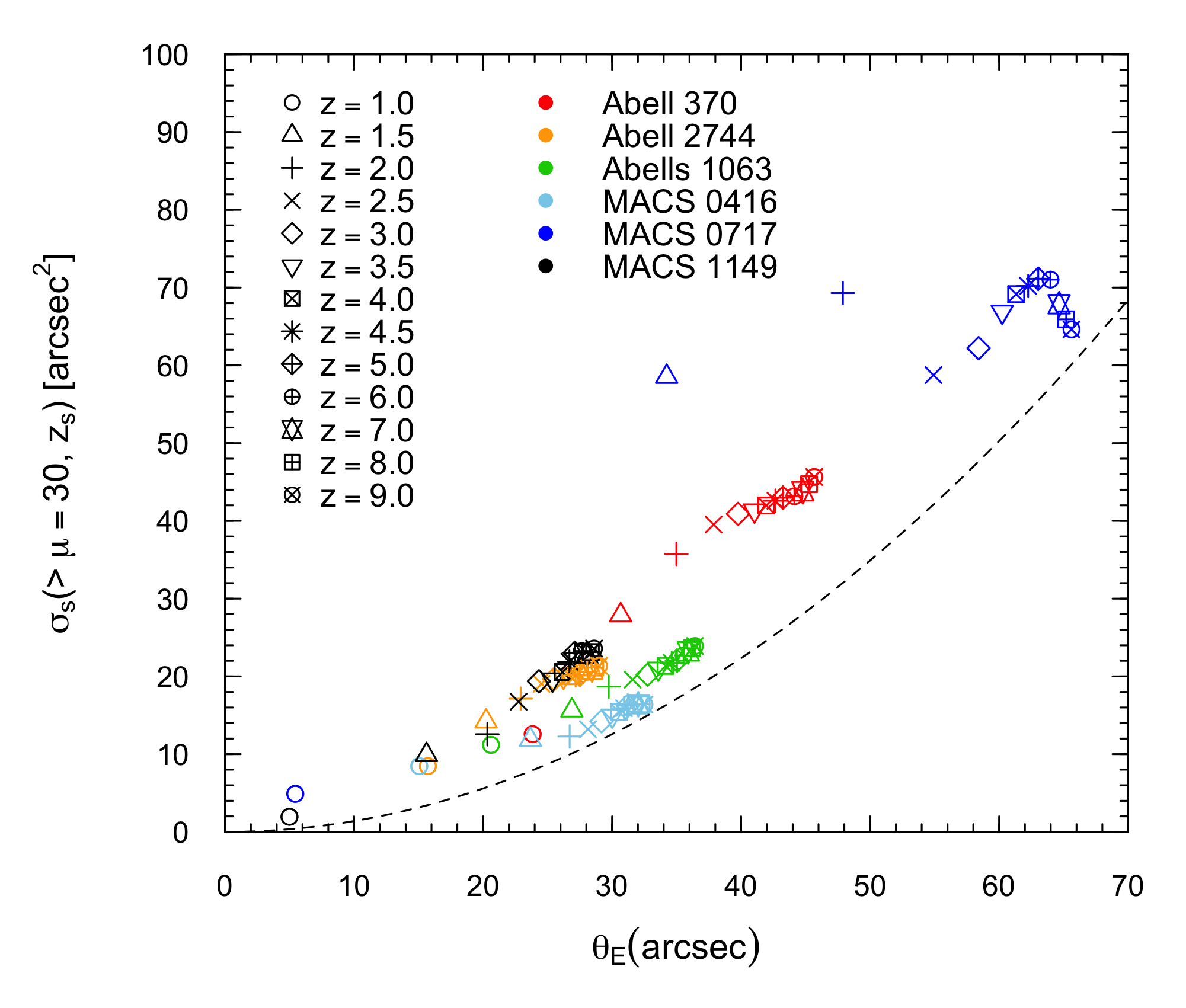}
 \caption{Area in the source plane for $\mu>30$ (in arcsec$^2$) as a function of the size of the Einstein radius (in arcsec). Different data points indicate the source redshift ($z_s$) at which $\sigma_s(>\mu=30)$ and $\theta_E$ are computed, while color coding denotes each HFF cluster. Data points are shown for different source redshifts between $z_s=1.0$ and $z_S=9.0$. Dashed line indicates the approximation given by equation~\ref{eq:As_cum}: $\sigma_s(>\mu=30, z_s) = (4\pi / 30^2)~\theta_E(z_s)^2$.
  \label{fig:area_theta}}
\end{figure}

The area in the source plane above a given magnification, $\sigma_s(>\mu)$, varies for each cluster (mass and redshift), but also depends on the redshift of the background source ($z_s$). This dependence is not trivial, as one would naively expect from the intrinsic dependence of the Einstein radius with the inverse of the \textit{lensing effective distance}, $D^{-1}=D_{ls}/(D_l~D_s)$. Instead, as the redshift of the background source increase, the critical curves grow and can trace new substructure that can modify the lensing efficiency.

To examine the dependence of $\sigma_s (>\mu)$ with the source redshift, in figure~\ref{fig:A0_source}, we show $\sigma_s(>\mu=30)$ as a function of $z_s$. The normalization is chosen for $\mu > 30$, since the relation $\sigma_s(>\mu) \propto \mu^{-2}$ is strictly valid only for large magnification factors. Note that the values of $\sigma_s(>\mu=30)$ for $z_s = 9$ are shown in table~\ref{tb:As}. Given the fact that the area in the source plane depends on both the redshift of the lens and the redshift of the source, we include in the figure the expected values of the inverse of the \textit{lensing effective distance} ($D^{-1}$) normalized by the value of $\sigma_s(>\mu=30, z_s = 3)$.  The values of $\sigma_s(>\mu=30, z_s)$ grows as $z_s$ increases following, to first order, the shape of $D^{-1}$. However, at redshifts $z_s\lesssim2$, it departs from the functional form of $D^{-1}$, probably due to the role of substructure. It is interesting to note that in the case of MACS 0717 the values of $\sigma_s(>\mu=30, z_s)$ decrease for $z_s \geqslant 2$. This is an artifact due to the limited size of the field of view, which is insufficiently large to include all the supercritical regions as can be clearly seen in the corresponding panels for MACS 0717 in figure~\ref{fig:causmaps}.

Finally, figure~\ref{fig:area_theta} shows the area in the source plane for $\mu>30$ as a function of the size of the Einstein radius, $\theta_E (z_s)$. For a circularly symmetric lens, one would expect a scaling relations like the one described in  equation~\ref{eq:As_cum}. Although this might be a good first order approximation, the structure of the magnification in the source plane is more complex and departures from this relation are clearly appreciated. Surprisingly, the dispersion in $\sigma_s(>\mu=30, z_s)$ for a given $\theta_E (z_s)$ is relatively small for the model of the HFF clusters here presented.

\section{Predicted number of lensed high-redshift galaxies}
\label{sec:lensedgalaxies}

Galaxy clusters can act as cosmic telescopes, magnifying the apparent bright and size of background galaxies, enabling the observation of distant sources which would be impossible to detect otherwise. Future observations with the JWST will reveal numerous distant faint galaxies. Thanks to the magnification boost provided by galaxy clusters, the faintest galaxies are expected to be found around gravitational lenses. In this section, we use the probability of magnification presented in the previous section to estimate the number of observed galaxies at a given magnitude in the field of view of the six HFF clusters. 

Given a classical luminosity function $\phi(M)$, in units of galaxies per absolute magnitude $M$ per unit volume in a given redshift interval d$z$, the lensed luminosity function, denoted by $\phi^*(M)$, in the same redshift interval is given by
\begin{equation}
\phi^*(M,z) = \frac{1}{d\Omega}\int_{\mu_{\rm{min}}}^{\mu_{\rm{max}}} \phi(M + 2.5\textrm{log}\mu,z)~\frac{\rm{d}\sigma_s (\mu,z)}{\rm{d}\mu}~\rm{d}\mu,
\end{equation}
where $\rm{d}\sigma_s (\mu,z)/\rm{d}\mu$ is the area in the source plane that is lensed with magnification between $\mu$ and $\mu+$d$\mu$, and at redshift $z$. The expected number of lensed galaxies is normalized by the corresponding total field of view (d$\Omega$, see table~\ref{tb:numarcs}) of each cluster.
The integral is computed between magnification $\mu_{\rm{min}}=1$ and $\mu_{\rm{max}}=100$. Magnification factors larger than 100 are not possible for galaxies larger than small dwarfs. On the lower limit, we ignore the effect of demagnification ($\mu<1$) since those galaxies are very unlikely to be observed at high $z$, and the field of view we have considered does not extend farther enough from the centres of the HFF clusters, where values of $\mu<1$ can be found. 

\begin{figure}
\centering
 \includegraphics[width=\columnwidth]{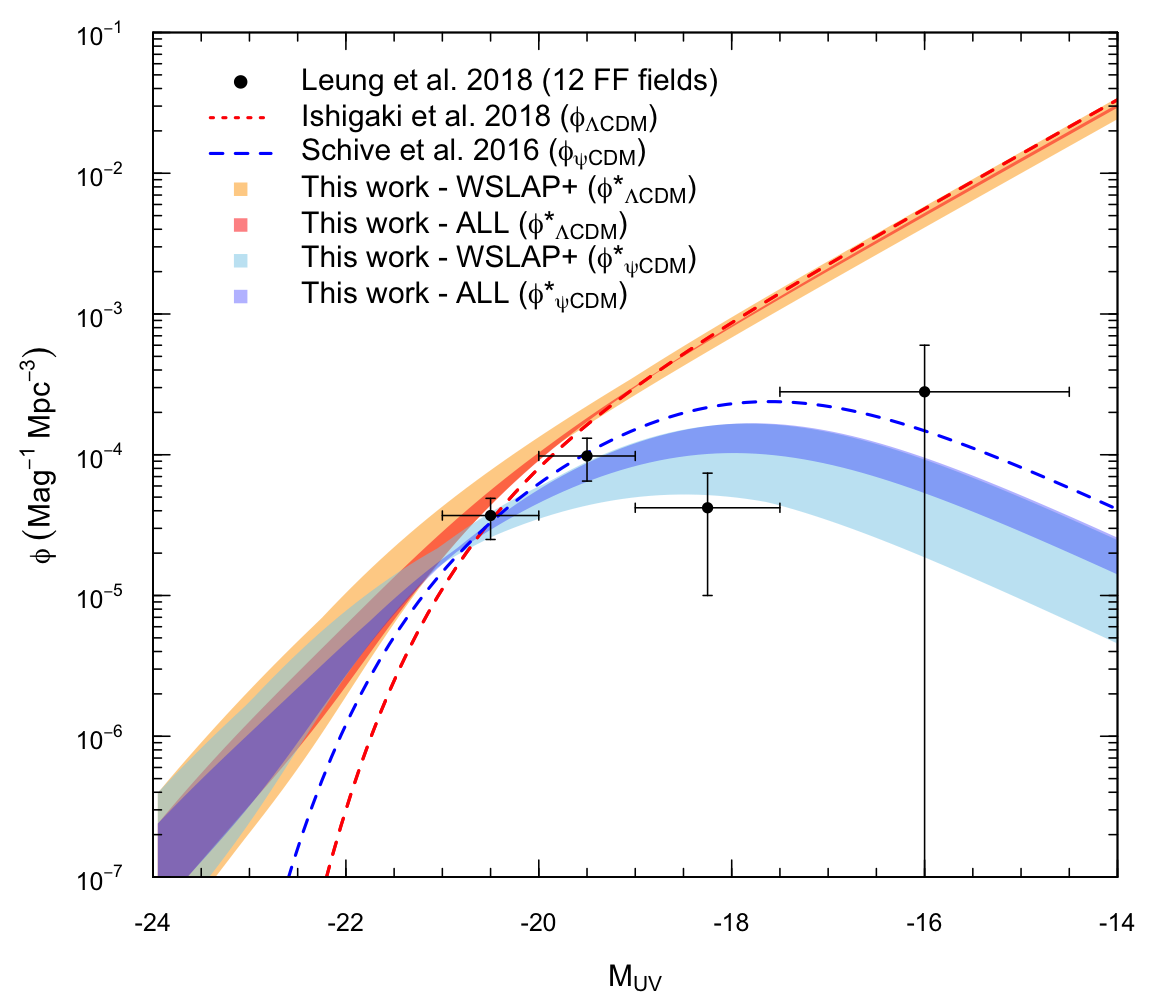}
 \caption{UV Luminosity Function at $z=9$ with $\Delta z=1.0$. The dashed red line corresponds to the UV LF described by a Schechter function for the $\Lambda$CDM model with the best-fit parameters derived by \citet{Ishigaki2018}, while the dashed blue line indicates the UV LF for Wave Dark Matter model ($\psi$DM) proposed by \citet{Schive2016} and based on the best-fit Schechter parameters found by \citet{Bouwens2015}. The dark orange and dark blue shaded regions correspond to the lensed UV LF by the six HFF clusters for $\Lambda$CDM and $\psi$DM, respectively, for the magnification models derived with \wslap. The light orange and light blue shaded regions correspond to the UV LF by the six HFF clusters for $\Lambda$CDM and $\psi$DM, respectively, for all the magnification models produced within the \textit{Frontier Fields Lensing Models v.4} collaboration. The black data points correspond to the results obtained by \citet{Leung2018} using data from the twelve HFF fields (clusters + parallels fields).
  \label{fig:uvlf}}
\end{figure}

The UV LF in the $\Lambda$CDM scenario is assumed to be described by a Schechter function \citep{Schechter1976}. We also consider deviations of the UV LF from the classical Schechter form. In particular, we consider also the alternative model of wave dark matter ($\psi$DM, \citealt{Schive2014}), which produces a suppression of small-scale structure. The UV LF of this model can be described as a modified Schechter function \citep[see][for more details]{Leung2018}.

In figure~\ref{fig:uvlf}, we show the lensing effects produced by the six HFF clusters on the observed UV LF of galaxies, denoted as $\phi^*(M,z)$, at $z=9$. For the UV LF of the $\Lambda$CDM, we use the best-fit Schechter parameters derived by \citet{Ishigaki2018}. For the $\psi$DM, we include the case with a mass for the DM bosons of $m_\textrm{B} = 0.8 \times 10^{-22}$eV, which corresponds to the model in \citet{Schive2016} with the strongest suppression of small-scale structures and, therefore, it may be consider as a lower limit for the UV LF in the $\psi$DM cosmological model. Lensing can lead to a modification of the number of galaxies at a given absolute magnitude in different ways. On one hand, galaxies that would have been detected even without lensing are brightened and, therefore, shifted towards smaller magnitudes depending on the magnification factor. Then, lensing leads to an excess (deficit) in galaxies brighter (fainter) than M$_{\textrm{UV}} \sim -20$. It is important to note that the given UV LF at the bright end are not exactly identical for $\Lambda$CDM and $\psi$DM, given that the former is the result of the recent study presented by \citet{Ishigaki2018}, while the later was proposed by \citet{Schive2016} based on the best-fit Schechter parameters found by \citet{Bouwens2015}. On the other hand, lensing prompts the flux of some galaxies (that could not otherwise be detected) above the detection threshold. 

The modification of the number of detectable galaxies due to gravitational lensing, i.e., $\phi^*(M,z)$, depends on the faint-end slope of the UV LF, which is the main difference between the $\Lambda$CDM and the $\psi$DM models. In the $\Lambda$CDM model, the approximately magnification-invariant faint-end slope of $\alpha \sim -2$ leads to number densities of galaxies towards fainter apparent magnitudes at the same rate as in the unlensed case \citep[see discussion in section 4 of][]{Leung2018}. However, in the $\psi$DM model, the presence of a faint-end turnover in the UV LF provides too few faint galaxies to compensate for the loss of galaxies being magnified and, hence, brightened towards lower absolute magnitudes. The light blue and light orange shaded region in figure~\ref{fig:uvlf} show the uncertainty in the lensed predictions when all public models are taken into account. Note how our models (dark blue shaded region) tend to predict more sources in the faint end. This is a consequence of the WSLAP+ models being biased low in terms of area at low magnifications $\mu\approx$1--2 (as shown by the area ratios in figure~\ref{fig:As_teams_ratio}). At these low magnifications, our models promote less sources to brighter fluxes leaving the lensed luminosity function closer to the underlying one.

Data points in figure~\ref{fig:uvlf} correspond to the recent results of \citet{Leung2018} for the UV LF using all the available data from the twelve HFF fields (six cluster and six parallel fields) from $z > 4.75$ to $z \sim 10$. For high absolute magnitudes, the results of \citet{Leung2018} are consistent with a slow rollover at the faint end of the UV LF that indicates a preference for Bose-Einstein condensate dark matter with a light boson mass (of the order of $m_\textrm{B} \simeq 10^{-22}$eV) over standard CDM cosmological model. In contrast, \cite{Ishigaki2018} derives a steeper luminosity function in the faint end of the luminosity function at $z\approx 9$, more in agreement with the standard Schechter function. Future data at $z>9$ obtained with the JWST will be able to settle this important question and discriminate between alternative models.

\section{Conclusions}
\label{sec:conclusions}
We present the gravitational lensing models for six galaxy clusters observed under the umbrella of the HFF program. The HFF clusters are powerful lenses at redshift $0.3 \lesssim z \lesssim 0.55$ observed with the HST to a depth of $m_{AB} = 29$ in the $r$-band. Each cluster contains of the order of one hundred multiple lensed images of distant background galaxies identified in HST color images by seven independent teams (including ours) within the \textit{Frontier Fields Lensing Models v.4} program. Many of these multiply lensed images have been reliably identified thanks to spectroscopic information provided by HST's GRISM instrument and VLT's MUSE instrument among others.

For the lensing reconstruction we use the free-form code \wslap~code based on available strong lensing data. The gravitational lensing maps presented in this paper allow us to draw the following conclusions:

\begin{itemize}

\item rms of the difference between the predicted and observed positions of the multiple-image systems used as constrains is below 1 arcsec for all the six HFF lens models derived with \wslap~code;
\item mapping the frequency of the values of the convergence and shear of a given lens model (what we called {\it lens' ghost}) and the location of the arcs in the same space ($\kappa-\gamma$) could be used as a self-consistency check of the lens models;

\item we derive magnification maps in the lens and the source plane for the six HFF clusters. MACS 0717 is the cluster with the most complex structure and also the one with the largest tangential critical curve ($\theta_E \approx 66$'' for $z_s = 9.0$); the sizes of the tangential critical curves (i.e., the Einstein radius) for the six HFF clusters are shown in table~\ref{tb:rms};

\item we compute the probability of lensing in terms of the area in the source plane above a given magnification, $\sigma_s (>\mu)$, using two different approaches: the so-called \textit{image plane} and \textit{source plane} methods. The \textit{image plane} method accounts for the magnification of a single counter-image (when it is the only one being observed), while the \textit{source plane} method accounts the magnification of all the counter-images of the same background source (when the counter-images are not individually resolved);

\item we derive the ratio of the \textit{source plane} to the \textit{image plane} methods for $\sigma_s (>\mu)$, denoted as $\sigma^s_s / \sigma^i_s$ for $z_s =$ 2, 3 and 9.  By definition, these ratios are below $\approx 3.0$ for any given $\mu$ and $z_s$, and above a factor of 2 for high magnification values ($\mu \gtrsim 20$). Table~\ref{tb:As} shows the values $\sigma_s (>\mu=30)$ for both methods along with the values of their ratios, labelled as $\sigma^s_s / \sigma^i_s$, at $\mu>30$ and for $z_s = 9.0$;

\item we check the consistency of our lens model by comparing them with the expectation of the lens models generated by the rest of the teams involved in the \textit{Frontier Fields Lensing Models v.4} program. We find a good agreement between models at large magnification factors, specially for the two clusters that are the most relaxed ones, AS1063 and MACS 1149. The models for  Abell 2744, MACS 0416 and MACS 0717 are the ones showing larger dispersion between the different teams. At low magnifications ($\mu \lesssim 3$), the differences are typically larger, up to a factor 3-4; 

\item when comparing our results with lens models derived within previous campaigns of the \textit{Frontier Fields Lensing Models}, the values of $\sigma_s (>\mu = 30)$ have increased by a factor of 1.4 to 2.3 depending on the cluster;

\item we show that the dependence of $\sigma_s (>\mu)$ with the source redshift ($z_s$) is expected to be proportional to $\theta_E^2$ for a circularly symmetric lens. Although this is just a first order approximation because the structure of the magnification in the source plane is more complex and departures from this first order approximation, surprisingly, the dispersion in $\sigma_s (>\mu=30, z_s)$ for a given $\theta_E (z_S)$ is relatively small for all the HFF lens models here presented;

\item we use the probability of magnification to estimate the number of observed galaxies at a given magnitude in the field of view of the six HFF clusters. We derive the lensed UV luminosity function, in units of galaxies per absolute magnitude per unit volume, for two different cosmological models: the $\Lambda$CDM scenario described by a Schechter function \citep{Schechter1976} and the wave dark matter model ($\psi$DM) described in \citealt{Schive2014} with a small-scale structure suppression.

\end{itemize}

\section{acknowledgements}
J.V-F and G.M.B. acknowledge support from the Space Telescope Science Institute (contract number 49726). J.M.D. acknowledges the support of projects AYA2015-64508-P (MINECO/FEDER, UE) funded by the Ministerio de Econom\'ia y Competitividad. J.V-F and J.M.D. acknowledge the hospitality of the University of Pennsylvania.
\bibliographystyle{mnras}
\bibliography{muFF}

\appendix
\section{Compilation of  multiple-images positions}
\label{app}
This appendix presents the sample of secure and likely lensed multiple images detected behind the six HFF clusters using the updated imaging from the \textit{Hubble Frontier Fields} program \citep{Lotz2017}, and spectroscopic redshifts from GLASS, CLASH-VLT, VLT/MUSE instrument and the literature. The multiple images are defined in common with other teams from the \textit{Frontier Fields Lensing Models v.4} collaboration.

Tables~\ref{tb:A370_arcs},\ref{tb:A2744_arcs},\ref{tb:AS1063_arcs},\ref{tb:M0416_arcs},\ref{tb:M0717_arcs} and~\ref{tb:M1149_arcs} in the printed version only show the first $\sim$10 rows. The full tables are available as online material in both ASCII and pdf formats.


\begin{table*}
  \caption{Abell 370 full strong lensing data set. The first column shows system ID (ID1.ID2.ID3 = System.Image.Knot). Second and third columns correspond to the system coordinates (in degrees). Column 4 indicates the redshift used for the lensing models. Column 5 show the system rank: \textsc{gold}, \textsc{silver} and \textsc{bronze}. Column 6 indicates the references from which the redshifts are taken (L17: \citealt{Lagattuta2017}; D18: \citealt{Diego2018c}; L-FFLens: D. Lagattuta within the \textit{Frontier Fields Lensing Models v.4} collaboration).}
 \label{tb:A370_arcs}
 \begin{tabular}{|cccccc} 
 \hline
 KnotID & RA & DEC & z & Rank & Comments\\
 \hline
1.1.1 & 39.967083 & -1.5769056 & 0.8041 & \textsc{gold} & L17 \\
1.2.1 & 39.976292 & -1.5760417 & 0.8041 & \textsc{gold} & L17 \\
1.3.1 & 39.968683 & -1.5765972 & 0.8041 & \textsc{gold} & L17 \\
\hline
2.1.1 & 39.973850 & -1.5842250 & 0.7251 & \textsc{gold} & L17 \\
2.2.1 & 39.970954 & -1.5850472 & 0.7251 & \textsc{gold} & L17 \\
2.3.1 & 39.968746 & -1.5845194 & 0.7251 & \textsc{gold} & L17 \\
2.4.1 & 39.969425 & -1.5847333 & 0.7251 & \textsc{gold} & L17 \\
2.5.1 & 39.969646 & -1.5848417 & 0.7251 & \textsc{gold} & L17 \\
\hline
3.1.1 & 39.965650 & -1.5668556 & 1.9553 & \textsc{gold} & L-FFLens \\
3.2.1 & 39.968529 & -1.5658111 & 1.9553 & \textsc{gold} & L-FFLens \\
...\\

 \end{tabular}  
\end{table*}

\begin{table*}
  \caption{Abell 2744 full strong lensing data set. The first column shows system ID (ID1.ID2.ID3 = System.Image.Knot). Second and third columns correspond to the system coordinates (in sexagesimal units). Column 4 indicates the redshift used for the lensing models. Column 5 show the system rank: \textsc{gold}, \textsc{silver} and \textsc{bronze}. Column 6 indicates the references from which the redshift are extracted (R14: \citealt{Richard2014}; Jh14: \citealt{Johnson2014}; J15: \citealt{Jauzac2015}; W15: \citealt{Wang2015}; M18: \citealt{Mahler2018}) or \textsc{knot} when there is a knot within the system.}
 \label{tb:A2744_arcs}
 \begin{tabular}{|cccccc} 
 \hline
 KnotID & RA & DEC & z & Rank & Comments\\
 \hline
 1 1 1  &  00 14 23.415  &     -30 24 14.15     & 1.6880 &     \textsc{gold}    &  M18\\
 1 1 2  &  00 14 23.296  &     -30 24 17.01     & 1.6880 &      \textsc{bronze}  &  M18\\
 1 1 3  &  00 14 23.382  &     -30 24 15.83     & 1.6880 &      --    &  \textsc{knot}\\
 1 1 4  &  00 14 23.458  &     -30 24 12.78     & 1.6880 &      --    &  \textsc{knot}\\
 1 1 5  &  00 14 23.321  &     -30 24 14.95     & 1.6880 &      --    &  \textsc{knot}\\
 1 2 1  &  00 14 23.032  &     -30 24 24.50     & 1.6880 &      \textsc{gold}    &  M18\\
 1 2 2  &  00 14 23.133  &     -30 24 22.14     & 1.6880 &      \textsc{bronze}  &  M18\\
 1 2 3  &  00 14 23.107  &     -30 24 23.77     & 1.6880 &      --    &  \textsc{knot}\\
 1 2 4  &  00 14 22.922  &     -30 24 25.98     & 1.6880 &      --    &  \textsc{knot}\\
 1 3 1  &  00 14 20.696  &     -30 24 35.99     & 1.6880 &      \textsc{gold}    &  M18\\
...\\
\end{tabular}
\end{table*}


\begin{table*}
  \caption{Abells 1063 full strong lensing data set. The first column shows system ID (ID1.ID2.ID3 = System.Image.Knot). Second and third columns correspond to the system coordinates (in degrees). Column 4 indicates the redshift used for the lensing models. Column 5 show the system rank: \textsc{gold}, \textsc{silver} and \textsc{bronze}. Column 6 indicates the references from which the  redshift are taken (B13: \citealt{Balestra2013}; Bo13: \citealt{Boone2013}; R14: \citealt{Richard2014}; Jh14: \citealt{Johnson2014}; K15: \citealt{Karman2015}; C16a: \citealt{Caminha2016a}; C16b: \citealt{Caminha2016b}; D16: \citealt{Diego2016c}; V16: \citealt{Vanzella2016}; K17: \citealt{Karman2017}; Cl-FFLens: B. Cl\'ement within the \textit{Frontier Fields Lensing Models v.4} collaboration) or \textsc{knot} when there is a knot within the system.}
 \label{tb:AS1063_arcs}
 \begin{tabular}{|cccccc} 
 \hline
 KnotID & RA & DEC & z & Rank & Comments\\
 \hline
1 1 1          &            342.194450    &     -44.527003     &       1.2278     &       \textsc{gold}     &       B13, C16a (K17 2b)\\
1 2 1          &            342.195867    &     -44.528950     &       1.2278     &       \textsc{gold}     &       B13, C16a (K17 2a)\\
1 3 1          &            342.186421    &     -44.521203     &       1.2278     &       \textsc{gold}     &       B13, C16a (K17 2c)\\
 \hline
2 1 1          &            342.192708    &     -44.531189     &       1.2593     &       \textsc{gold}     &       B13, C16a (K17 3a)       \\
2 2 1          &            342.192125    &     -44.529831     &       1.2593     &       \textsc{gold}     &       B13, C16a (K17 3b)\\
2 3 1          &            342.179863    &     -44.521561     &       1.2593     &       \textsc{gold}     &       B13, R14, J14, C16a (K17 3c)\\
 \hline
3 1 1          &            342.195542    &     -44.532139     &       1.7000     &       \textsc{silver}     &       D16\\
3 2 1          &            342.193917    &     -44.528731     &       1.7000     &       \textsc{silver}     &       D16\\
 \hline
4 1 1          &            342.193708    &     -44.530161     &       1.2583     &       \textsc{gold}     &       K17 (13a)\\
...\\
 \end{tabular}
\end{table*}


\begin{table*}
  \caption{MACS 0416 full strong lensing data set. The first column shows system ID (ID1.ID2.ID3 = System.Image.Knot). Second and third columns correspond to the system coordinates (in degrees). Column 4 indicates the redshift used for the lensing models. Column 5 show the system rank: \textsc{gold}, \textsc{silver} and \textsc{bronze}. Column 6 indicates the references from which the  redshift are taken (J14: \citealt{Jauzac2014}; C17: \citealt{Caminha2017}) or \textsc{knot} when there is a knot within the system.}
 \label{tb:M0416_arcs}
 \begin{tabular}{|cccccc} 
 \hline
 KnotID & RA & DEC & z & Rank & Comments\\
 \hline
  1 1 1   &   64.04075   &  -24.06159  &  1.8960   &  \textsc{gold}  &  J14 (1.1) \\
  1 1 2   &   64.04117   &  -24.06186  &  1.8960   &  \textsc{gold}  &  J14 (2.1) \\
  1 2 1   &   64.04348   &  -24.06354  &  1.8960   &  \textsc{gold}  &  J14 (1.2) \\
  1 2 2   &   64.04302   &  -24.06302  &  1.8960   &  \textsc{gold}  &  J14 (2.2) \\
  1 3 1   &   64.04735   &  -24.06867  &  1.8960   &  \textsc{gold}  &  J14 (1.3) \\
  1 3 2   &   64.04746   &  -24.06883  &  1.8960   &  \textsc{gold}  &  J14 (2.3) \\
 \hline
  2 1 1   &   64.03077   &  -24.06712  &  1.9900   &  \textsc{gold}  &  J14 (3.1) \\
  2 1 2   &   64.03077   &  -24.06722  &  1.9900   &  \textsc{gold}  &  J14 (4.1) \\
  2 1 3   &   64.03099   &  -24.06731  &  1.9900   &  --                   &  \textsc{knot} \\
  2 2 1   &   64.03525   &  -24.07098  &  1.9900   &  \textsc{gold}  &  J14 (3.2) \\
 ...\\
  \end{tabular}
\end{table*}


\begin{table*}
  \caption{MACS 0717 full strong lensing data set. The first column shows system ID (ID1.ID2.ID3 = System.Image.Knot). Second and third columns correspond to the system coordinates (in degrees). Column 4 indicates the redshift used for the lensing models. Column 5 show the system rank: \textsc{gold}, \textsc{silver} and \textsc{bronze}. Column 6 indicates the references from which the  redshift are taken (K16: \citealt{Kawamata2016}; L16: \citealt{Limousin2016}) or \textsc{knot} when there is a knot within the system.}
 \label{tb:M0717_arcs}
 \begin{tabular}{|cccccc} 
 \hline
 KnotID & RA & DEC & z & Rank & Comments\\
 \hline
1 1 1   &   07 17 34.881   &    +37 44 28.23   &    2.9630   &    \textsc{gold}   &   K16 (1.1) \\
1 1 2   &   07 17 34.874   &    +37 44 28.04   &    2.9630   &    --                     &   \textsc{knot} \\
1 1 3   &   07 17 34.918   &    +37 44 29.03   &    2.9630   &    --                     &   \textsc{knot} \\
1 1 4   &   07 17 34.910   &    +37 44 28.65   &    2.9630   &    --                     &   \textsc{knot} \\
1 1 5   &   07 17 34.762   &    +37 44 26.27   &    2.9630   &    --                     &   \textsc{knot} \\
1 1 6   &   07 17 34.792   &    +37 44 26.19   &    2.9630   &    --                     &   \textsc{knot} \\
1 1 7   &   07 17 34.962   &    +37 44 28.56   &    2.9630   &    --                     &   \textsc{knot} \\
1 1 8   &   07 17 34.964   &    +37 44 30.20   &    2.9630   &    --                     &   \textsc{knot} \\
1 2 1   &   07 17 34.518   &    +37 44 24.33   &    2.9630   &    \textsc{gold}   &   K16 (1.2) \\
1 2 2   &   07 17 34.534   &    +37 44 24.44   &    2.9630   &    --                     &   \textsc{knot} \\
...\\
 \end{tabular}
\end{table*}


\begin{table*}
  \caption{MACS 1149 full strong lensing data set. The first column shows system ID (ID1.ID2.ID3 = System.Image.Knot). Second and third columns correspond to the system coordinates (in degrees). Column 4 indicates the redshift used for the lensing models. Column 5 show the system rank: \textsc{gold}, \textsc{silver} and \textsc{bronze}. Column 6 indicates the references from which the  redshift are taken (J16: \citealt{Jauzac2016}) or \textsc{knot} when there is a knot within the system.}
 \label{tb:M1149_arcs}
 \begin{tabular}{|cccccc} 
 \hline
 KnotID & RA & DEC & z & Rank & Comments\\
 \hline

 1 1  1   &   11 49 35.282   &   +22 23 45.64   &   1.4880   &  Gold  &   J16 \\
 1 1  2   &   11 49 35.213   &   +22 23 43.35   &   1.4880   &  --    &  \textsc{knot}    \\
 1 1  3   &   11 49 35.575   &   +22 23 44.27   &   1.4880   &  --    &  \textsc{knot} (SN Refsdal) \\
 1 1  3   &   11 49 35.453   &   +22 23 44.82   &   1.4880   &  --    &  \textsc{knot} (SN Refsdal) \\
 1 1  3   &   11 49 35.369   &   +22 23 43.94   &   1.4880   &  --    &  \textsc{knot} (SN Refsdal) \\
 1 1  3   &   11 49 35.474   &   +22 23 42.68   &   1.4880   &  --    &  \textsc{knot} (SN Refsdal) \\
 1 1  4   &   11 49 35.158   &   +22 23 44.16   &   1.4880   &  --    &  \textsc{knot}    \\
 1 1  5   &   11 49 35.558   &   +22 23 46.86   &   1.4880   &  --    &  \textsc{knot}    \\
 1 1  6   &   11 49 35.383   &   +22 23 47.09   &   1.4880   &  --    &  \textsc{knot}    \\
 1 1  7   &   11 49 35.306   &   +22 23 48.19   &   1.4880   &  --    &  \textsc{knot}    \\
 ...\\
 \end{tabular}
\end{table*}

\label{lastpage}
\end{document}